\documentclass[iop]{emulateapj}

\def\msun{\hbox{M$_{\odot}$}}
\def\mdot{\hbox{$\dot M$}}
\def\micron{$\mu$m}
\def\microns{$\mu$m}

\newcommand\be{\begin{equation}}
\newcommand\en{\end{equation}}

\begin{document}

\shortauthors{Espaillat et al.}
\shorttitle{IR Disk Variability}

\slugcomment{Accepted to ApJ: December 7, 2010}

\title{A {\it Spitzer} IRS Study of Infrared Variability in Transitional and 
Pre-Transitional Disks around T Tauri Stars}

\author{
C. Espaillat\altaffilmark{1,2}, 
E. Furlan\altaffilmark{3,4}, 
P. D'Alessio\altaffilmark{5},
B. Sargent\altaffilmark{6},
E. Nagel\altaffilmark{7}, 
N. Calvet\altaffilmark{8}, 
Dan M. Watson\altaffilmark{9},
J. Muzerolle\altaffilmark{6}
}

\altaffiltext{1}{NSF Astronomy \& Astrophysics Postdoctoral Fellow}
\altaffiltext{2}{Harvard-Smithsonian Center for Astrophysics, 60 Garden
Street, MS-78, Cambridge, MA, 02138, USA; cespaillat@cfa.harvard.edu}
\altaffiltext{3}{{\it Spitzer} Fellow} 
\altaffiltext{4}{Jet Propulsion Laboratory, California Institute of Technology, Mail 
Stop 264-767, 4800 Oak Grove Drive, Pasadena, CA 91109, USA; Elise.Furlan@jpl.nasa.gov} 
\altaffiltext{5}{Centro de Radioastronom\'{i}a y Astrof\'{i}sica,
Universidad Nacional Aut\'{o}noma de M\'{e}xico, 58089 Morelia,
Michoac\'{a}n, M\'{e}xico; p.dalessio@crya.unam.mx}
\altaffiltext{6}{Space Telescope Institute, 3700 San Martin Drive,
Baltimore, MD 21218, USA; muzerol@stsci.edu}
\altaffiltext{7}{Departamento
de Astronom\'\i a, Universidad de Guanajuato, Guanajuato, Gto, M\'exico
36240; erick@astro.ugto.mx} 
\altaffiltext{8}{Department of
Astronomy, University of Michigan, 830 Dennison Building, 500 Church
Street, Ann Arbor, MI 48109, USA; ncalvet@umich.edu}
\altaffiltext{9}{Department of Physics and
Astronomy, University of Rochester, NY 14627-0171, USA;
dmw@pas.rochester.edu}

\begin{abstract}

We present a {\it Spitzer} IRS study of variability in 14
T Tauri stars in the Taurus and Chamaeleon star-forming regions. The
sample is composed of transitional and pre-transitional objects which
contain holes and gaps in their disks. We detect
variability between 5--38~{\micron} in all but two of our objects on timescales of 2--3 years. 
Most of the
variability observed can be classified as seesaw behavior, whereby the
emission at shorter wavelengths varies inversely with the emission at
longer wavelengths.   For many of
the objects we can reasonably reproduce the observed variability
using irradiated disk models, particularly by changing the
height of the inner disk wall by $\sim$20$\%$.  When the inner wall is taller, the
emission at the shorter wavelengths is higher since the inner wall
dominates the emission at 2--8~{\micron}.  The taller inner wall casts a
larger shadow on the outer disk wall, leading to less emission at
wavelengths beyond 20~{\micron} where the outer wall dominates.  
We discuss how the possible presence of planets in these disks could lead to warps which cause changes in the height of the inner wall. We also
find that
crystalline silicates are common in the outer disks of our objects and that in the four disks in the sample with the most crystalline silicates, variability on timescales of 1~week is present.
In addition to explaining the infrared
variability described above, planets can create shocks and collisions
which can crystallize the dust and lead to short timescale variability.

\end{abstract}


\section{Introduction} \label{intro}

T Tauri stars (TTS) are by definition variable stars, named after their
prototype T Tau. Variability in this class of objects is ubiquitous and
has inspired a long history of study \citep[e.g.,][]{joy45, joy49,
rydgren76}. The majority of more recent studies have focused on their
optical and near-infrared (NIR) emission. \citet{carpenter01} found
significant variability of TTS in Orion A based on near-IR JHK
2MASS photometry.  The typical timescale of variation in these TTS was
on the order of days and could be explained by cool spots, hot spots,
extinction, and$/$or changes in the mass accretion rate onto the star.
Work by \citet{eiroa02} further supported these results.  For many TTS
in that study the optical and JHK photometry varied simultaneously,
supporting that in most objects variability is due to star spots and
variable extinction.  However, for several of the objects in that sample
there is no correlation between the optical and NIR.  This points to
structural changes in the disk, which begin to dominate the SED in the
K-band, on the order of days.  

With the arrival of the {\it Spitzer Space Telescope} \citep{werner04},
more detailed studies probing the mid-IR have been possible,
particularly with the Infrared Spectrograph \citep[IRS;][]{houck04}
which provides simultaneous wavelength coverage between
$\sim$5 and 38~{\microns}. Variations in the shape and size of the
10~{\micron} silicate emission feature has been seen in DG Tau, XZ Tau,
\citep{bary09} and LRLL 31 \citep{muzerolle09}.  In the case of EX Lupi,
transient changes in the dust composition of the disk have been detected with
multi-epoch spectra of the silicate emission feature \citep{abraham09}.

Flux changes in the IRS spectra of disks have also been observed.
\citet{muzerolle09} found substantial variability in LRLL 31, located in
IC 348, on timescales down to days. The flux of this object oscillated
around a pivot point at 8.5~{\microns} -- as the emission decreased at
wavelengths shortwards of the pivot point, the emission increased at
longer wavelengths and vice versa. The star spots proposed to explain
variability at shorter wavelengths could change the irradiation heating, but
this would cause an overall increase or decrease of the flux, not an
anti-correlation between the flux centered at the pivot point.
\citet{muzerolle09} proposed that the observed ``seesaw'' variability
was due to dynamical changes in the disk itself. In particular, changes
in the height of the inner disk edge or wall located at the dust
destruction radius (which emits primarily in the NIR) could lead to
variable shadowing of the outer disk material (which emits at longer
wavelengths). Indeed, the overall characteristics of the flux changes
observed can be explained by models of a disk with an inner warp
that leads the scale height of the inner disk to change with time
\citep{flaherty10}.

It is important to note that the above-mentioned object, LRLL 13, is
surrounded by a transitional disk. Transitional disks have nearly
photospheric near-IR (1--5~{\micron}) and mid-IR (5--20~{\micron})
emission below the median excess of Class II objects, coupled with
substantial emission above the stellar photosphere at wavelengths beyond
$\sim$20~{\micron} \citep{strom89}.  This dip in the infrared flux has
been attributed to a central ``hole'' in the dust distribution of the
disk. This has been inferred from detailed modeling of some of these
transitional objects \citep{calvet02,calvet05,espaillat07a,espaillat08b}
and confirmed with sub-millimeter and millimeter interferometric imaging
\citep{hughes07,hughes09,andrews09}.

Motivated by the variability observed in the transitional disk of LRLL~13, 
we conducted a {\it Spitzer} IRS variability study of transitional
disks and pre-transitional disks. Pre-transitional disks have
significant near-infrared excesses relative to their stellar
photospheres, similar to the median spectral energy distribution of
disks in Taurus \citep{dalessio99}, while also exhibiting the
characteristics seen in transitional disks (i.e. deficits of
mid-infrared flux  and substantial excesses beyond $\sim$20~{\micron}).
This indicates a gapped disk structure where the inner disk is separated
from the outer disk. Sub-millimeter and millimeter interferometric
imaging \citep[Andrews et al, in prep;][]{andrews09, pietu06} has confirmed the location of the wall of the
outer disk inferred from SED modeling for a few pre-transitional disks
(e.g. LkCa~15, UX~Tau~A, Rox 44).  The near-IR
emission of these objects is due to dust in an optically thick inner disk, a result
obtained by using the ``veiling"
\citep{hartigan89} of near-infrared spectra
\citep{espaillat08a, espaillat10}.

Here we perform detailed modeling of the broad-band spectral energy
distributions of the 14 transitional and pre-transitional
disks in our sample at different epochs.  We take
into account the effect of shadowing by the inner disk on the outer disk
by employing the irradiated accretion disk models of \citet{dalessio06}
with the modifications to include shadowing presented in
\citet{espaillat10}. Our work adds to the number of detailed modeling
efforts of disk variability in the literature which have found some
success in reproducing the observations by varying the height of the
inner disk wall \citep{juhasz07, sitko08}. \citet{juhasz07}'s study of 
the UX Ori-type star SV Sep came to the conclusion that the variability
in the optical and near-IR emission could be explained by changing the
height of the inner disk edge, but they were unable to simultaneously
fit the variability from the IR out to 100~{\micron} using a
self-shadowed disk. \citet{sitko08} reported the variability of two
Herbig Ae stars between 1-5~{\micron}, the region dominated by the inner disk
wall, and could explain this change by varying the height of the inner
disk edge.  However, this work did not have data at longer wavelengths
and so could not test if these models fit the emission from the outer
disk.

The advantage of our study over previous works is the quality of
the data and the simultaneous wavelength coverage (5-38~{\micron})
provided by {\it Spitzer} IRS. Thus we are able to present the largest
and most detailed modeling study of variability in disks around TTS to
date.  We find that we can explain the variability of most of the
pre-transitional disks in the sample by
changing the height of the inner disk wall and thus the extent of its
shadow on the outer disk, thereby affecting the resulting emission from
the outer disk.  We also find that the objects in the sample with the
largest amounts of crystalline silicates in their disks exhibit variability on the
shortest timescales observed in this study.

\section{Sample Selection} \label{sample}

Our sample of 14 objects was chosen to include bright transitional and
pre-transitional disks in Taurus and Chamaeleon, two nearby 1--2 Myr old
star--forming regions with low extinction that were well covered by a
{\it Spitzer} guaranteed-time observing (GTO) program \citep[Manoj et
al., submitted;][]{furlan06, kim09}. For each of our 14 objects we obtained
a pair of general observer (GO) observations taken within 1 week of each
other.  When combined with the GTO data, these observations give us
baselines of $\sim$3--4 yr and $\sim$1 wk.

We included the well-studied and previously modeled transitional disks
CS~Cha, DM~Tau, and GM~Aur \citep{espaillat07a,calvet05} and the
pre-transitional disks LkCa~15 and UX~Tau~A \citep{espaillat07b}. We
also included other transitional and pre-transitional disks in
Chamaeleon (T25, T35, T56, SZ~Cha) which had been identified by
\citet{kim09}.  Four objects in our sample were chosen based on analysis
conducted in \citet{furlan09}.  In that work, we compared the observed
equivalent width of the 10~{\micron} silicate emission feature and the
SED slope between 13 and 31~{\microns} against a grid of disk models. 
RY~Tau, IP~Tau, CR~Cha, and WW~Cha were four of the many objects that
fell outside of range of EW(10~{\micron}) and n$_{13-31}$ covered by the
full disk model grid.  One explanation proposed by \citet{furlan09} and
\citet{espaillat09} to explain these outliers was that these disks are
actually pre-transitional disks with smaller gaps than had been
previously observed.  Therefore, we included these objects in the sample
based on their potential for being pre-transitional disks. ISO~52 is the
only object in our sample which could be explained by the full disk
model grid presented in \citet{furlan09}.  However, we chose to include
this object in our variability study because qualitatively its GTO IRS
spectrum resembled what would be expected from a pre-transitional disk
with a small gap \citep[see Figure 12 in][]{espaillat10}.
Except for CS~Cha, the objects in our
sample are thought to be single stars and so the holes and gaps in our
objects indicate the likely presence of planets 
\citep[see discussion in][]{espaillat08a,espaillat10}.\footnote{Even in the case of the spectroscopic
binary of CS~Cha, the 38~AU hole is too large to be cleared out by the
binary system alone.  A binary system with a circular orbit can clear a
hole twice the size of the semi-major axis; a binary with an
eccentricity of 0.8 can clear out a region 3.5 times the semi-major axis
\citep{artymowicz94,pichardo05,aguilar08}. Since the separation of the
binary in CS~Cha is $\sim$4~AU \citep{guenther07}, the binary should
clear only a region up to 14~AU.}

\section{Data Reduction} \label{redux}

\subsection{Observations}

Here we present three {\it Spitzer} IRS spectra for each of our targets
(Table~\ref{tab:log}). The first spectra were obtained
through IRS GTO in Program 2 (PI: Houck) and have been previously
presented elsewhere \citep[Manoj et al., submitted;][]{furlan06,furlan09,
kim09}.  The last two spectra for each target were obtained in GO
Program 50403 (PI: Calvet).  For consistency, we have re-reduced the GTO
data in the same way we reduce the GO data (see {\S}~\ref{datredux}).
We note that we also searched the {\it Spitzer} archive for all other IRS observations
of objects in our sample.  We reduced those data as we did
our GTO and GO data and comment further on these additional spectra in
the Appendix.

All of the GO observations were performed in staring mode using the
low-resolution modules (Short-Low (SL) and Long-Low (LL)) of IRS, spanning
wavelengths from 5--14~{\microns} and 14-38~{\microns}, respectively,
with a resolution $\lambda/\delta\lambda\sim$90. The Chamaeleon GTO
spectra were obtained in staring mode as well while the Taurus GTO
spectra were obtained in mapping mode with 2$\times$3 step maps (2
parallel and 3 perpendicular to the slit) on the target. Most of the GTO
spectra use the SLLL configuration.  The exceptions are RY~Tau, CR~Cha,
and WW~Cha which were taken with the SL module and the high-resolution
modules Short-High (SH) and Long-High (LH), which cover 10--19~{\micron}
and 19--37~{\micron}, with $\lambda/\delta\lambda\sim$600.
We note that in the case of DM~Tau, two GTO spectra were obtained in
Program 2 and are listed in Table~\ref{tab:log} as GTO1 and GTO2. 
Throughout this paper, we only show the GTO2 spectrum since it was taken
in staring mode, as were our GO observations.  In addition, the SL
spectrum has a higher SNR in the GTO2 observation due to a longer
integration time. We note that the GTO1 spectrum is equivalent to the
GTO2 spectrum (i.e. flux, shape) within the uncertainties of the
observations.

\subsection{Extraction and Calibration of Spectra} \label{datredux}

\begin{figure}
\epsscale{1}
\plotone{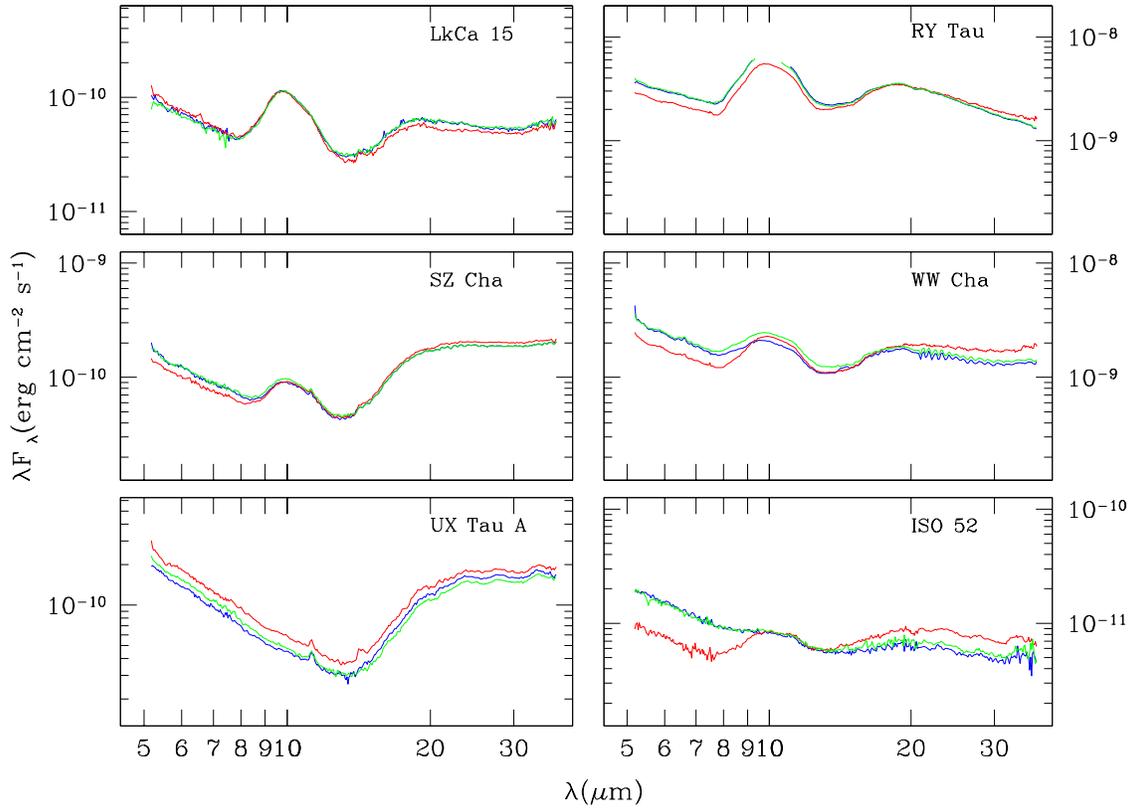}
\vskip -1.5in
\caption[]{{\it Spitzer} IRS spectra for pre-transitional disks in our sample which exhibit
seesaw-like behavior.
All epochs are shown:  GTO (red), GO1 (green), GO2 (blue).
Variability is apparent in all the objects.  None of the spectra have been corrected for extinction.
Note that for RY~Tau the region around 10~{\micron} in the GO spectra has been removed
since they were saturated in this region.  
The GTO observations of RY~Tau and the GO observations of WW Cha were significantly mispointed 
and so the variability seen here should be taken as an indication of the overall behavior 
of the variability in these objects and not as an accurate measurement of the amount of variability.  We note that the
fringing observed longward of 20~{\micron} in WW Cha is due to
mispointing.
(A color version of this figure is available in the online journal.)
}
\label{figirsptd1}
\end{figure}

\begin{figure}
\epsscale{1}
\plotone{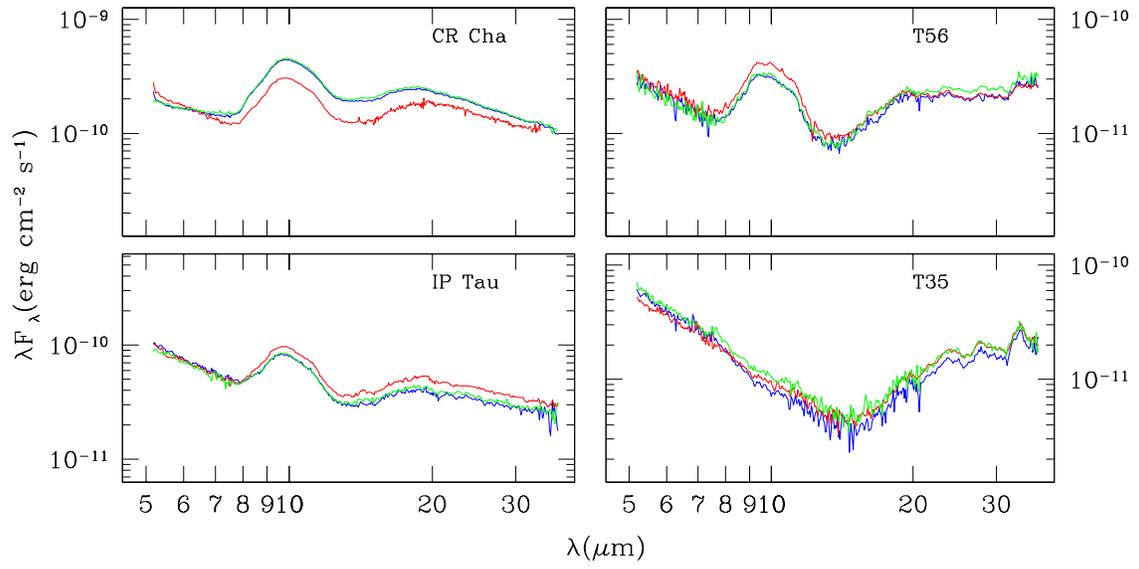}
\vskip -2.5in
\caption[]{{\it Spitzer} IRS spectra for pre-transitional disks in our sample where there is no clear
pivot point.
The color scheme is the same as that used in Figure~\ref{figirsptd1}.
(A color version of this figure is available in the online journal.)
}
\label{figirsptd2}
\end{figure}

\begin{figure}
\epsscale{1}
\plotone{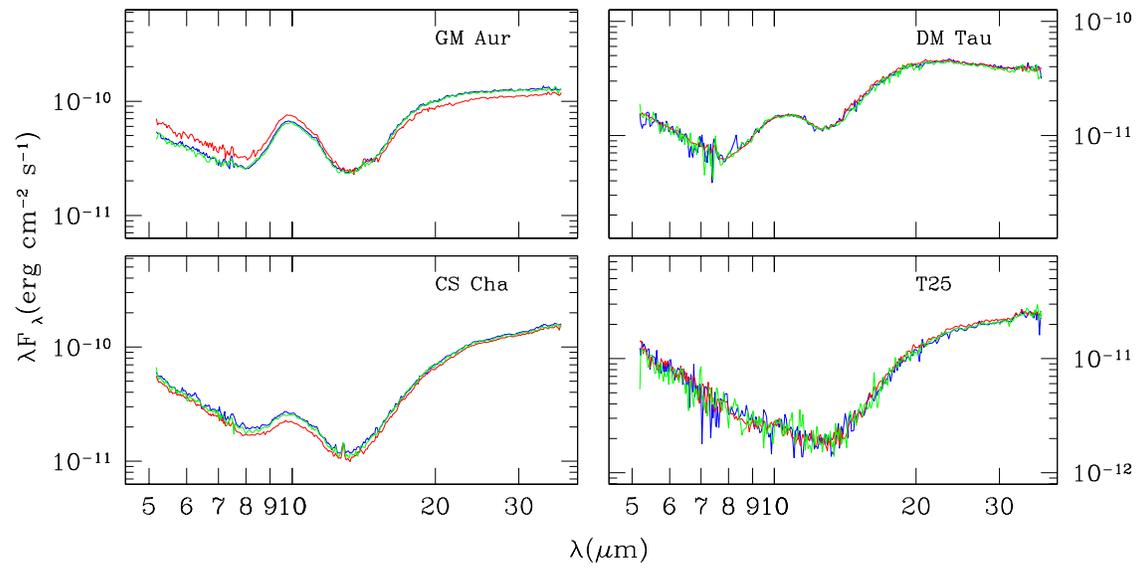}
\vskip -2.5in
\caption[]{{\it Spitzer} IRS spectra for the transitional disks in our sample.
The color scheme is the same as that used in Figure~\ref{figirsptd1}.
No variability is observed in DM~Tau and T25.
(A color version of this figure is available in the online journal.)
}
\label{figirstd}
\end{figure}

Details on the observational techniques and general data reduction
steps, including bad pixel identification, sky subtraction, and flux
calibration, can be found in \citet{furlan06} and  \citet{watson09}.
Here we provide a brief summary. Each object was observed twice along
the slit, at a third of the slit length from the top and bottom edges of
the slit. Basic calibrated data (BCD) with pipeline version S18.7 for
both the GTO and GO observations were obtained from the {\it Spitzer}
Science Center.  With the BCDs, we extracted and calibrated the spectra
using the SMART package \citep{higdon04}. Bad and rogue pixels were
corrected by interpolating from neighboring pixels. For the
low-resolution spectra, the data were sky subtracted using the sky
emission in the off-target nods except in the cases of CS~Cha GO1, IP
Tau GO 1, and LkCa~15 GO 2 where the off-target orders were used in
order to minimize over-subtraction of H~I from the sky background at
$\sim$20~{\micron}. For the GTO SH and LH modules, no background subtraction
was performed and so the emission of these targets could be slightly
overestimated, particularly in LH since the slit is larger. However, the
targets are much brighter than the background, and so the emission from
the disks should clearly dominate in the mid-IR. After sky subtraction,
the low-resolution spectra were extracted from the 2D spectral images
using a tapered column which varies with the width of the IRS
point-spread function.  For the SH and LH modules, a full slit 
extraction was performed. To flux calibrate the observations we used
spectra of $\alpha$ Lac (A1 V) for the low-resolution modules and $\xi$
Dra (K2III) for the high resolution modules.  We performed a nod-by-nod
division of the target spectra and the $\alpha$ Lac or $\xi$ Dra spectra
and then multiplied the result by a template spectrum \citep{cohen03}.
The final spectrum was produced by averaging the calibrated spectra from
the two nods. For the 2$\times$3 maps, only the central map positions
were used for the final spectrum. The high-resolution data were rebinned
to same sampling as the low-resolution data. Our spectrophotometric
accuracy is 2--5$\%$ estimated from half the difference between the
nodded observations. We note that there are artifacts in the T25 GO1,
ISO~52 GO1, T56 GO1, and CR~Cha GTO spectra at $\sim$9~{\micron},
$\sim$7~{\micron}, $\sim$15~{\micron}, and $>$35~{\micron},
respectively.  These spikes are due to additional bad pixels not
captured by the bad pixel and rogue masks used in our data reduction. 
For clarity, we manually mask these artifacts from the spectra. The
final spectra used in this study are shown in
Figures~\ref{figirsptd1},~\ref{figirsptd2},~and~\ref{figirstd}.

\subsection{Uncertainties due to Mispointing} \label{mispoint}

Apparent variability could also be the effect of mispointings, which
would cause loss in flux especially in SL, which is the module with the
narrowest slit width. PCRS peak-up, which we used for our GO
observations, yields a pointing accuracy of 0.4$^\prime$$^\prime$
(1-sigma radial rms); this is just slightly less than the accuracy for
blind pointing (0.5$^\prime$$^\prime$; however, note that earlier in the
mission the blind pointing of Spitzer was only accurate to
$<$1$^\prime$$^\prime$). According to a detailed study in
\citet{swain08}, a 0.4$^\prime$$^\prime$ pointing offset in SL1 causes a
drop in flux of about 7$\%$, with a slight (1-2$\%$) dependence on
wavelength. Therefore, if we observe a flux variation $\geq$10$\%$ that
fluctuates with wavelength more than on the few percent level, we can
most likely exclude mispointing as the cause of the variability.

The GTO observations did not use peak-ups.  However, the GTO data of our Chamaeleon
objects were taken later in the mission, when the blind pointing of {\it Spitzer}
was improved.
The objects in our sample that could be affected by some mispointing are
the Taurus GTO observations, taken in 2004 February. They were obtained
with blind pointing in mapping mode; the spectra presented here are
extracted from the central map positions. For IP Tau, DM Tau GTO1, and LkCa
15, the SL and LL spectra match at 14~{\micron}, which argues against
mispointing; since the SL and LL slits are perpendicular to each other,
mispointings should affect one module more than the other, resulting in
an offset. There were small (5$\%$) offsets between SL and LL for DM Tau GTO2, GM Aur
and UX Tau A, with SL lower than LL, suggesting small mispointings. RY
Tau likely suffered from larger mispointing, since SH is lower than SL
and LH by 20-30$\%$ (the mispointing relative to SL is more difficult to
determine, given that part of SL1 in the GO spectra is saturated); also,
compared to the GO data, the SL spectrum appears low. Overall, the GTO
spectrum of RY Tau is quite uncertain, but the other GTO observations
should have a pointing accuracy of 0.5$^\prime$$^\prime$ or better. In
order to account for the mispointing discussed here we scale the SL spectra
upward so that the SL and LL spectra match.  
We note that the high-resolution
modules SH and LH are much less affected by small mispointings,
since the slits are wider than for the low-resolution modules.

We note that none of the other GTO or GO observations were mispointed.
Aside from those observations discussed above, there is no mismatch between SL and LL, so we can be confident that
the observations are well pointed and that therefore both the SL and LL
slits contain the full flux of the object.
As an additional check, we find that the sources are located 
within $\sim$1$^\prime$$^\prime$ of each other
in the SL and LL slits amongst the GTO, GO1, and GO2 observations, further evidence
that these sources were well-centered in the slit.  
We note that at $\sim$14{\micron} the SL module ends and the LL module
begins.  Given that several of our objects have large holes and gaps, the emission of
the outer wall begins to dominate at 15--20{\microns}, coinciding with
where SL ends and LL starts.  

We also overlaid the IRS slit positions for each AOR on 2MASS K-band
images to check for anomalous behavior.  In WW~Cha GO1 and GO2 both of
the modules were mispointed due to an error in the coordinates (the
declination was off by $\sim$1$^\prime$$^\prime$). Judging from the overlay,  the LL
module was more off-target than the SL module and the mispointing
is more along the spatial direction of the SL slit, but it is difficult to
tell how much flux was lost. We note that in the GO observations of
SZ~Cha, two faint objects entered the SL and LL slits at
5.3$^\prime$$^\prime$ and 12.5$^\prime$$^\prime$ from the target.  However, it is known
that these two objects are not members of Cha I and are likely very
faint at mid-IR wavelengths \citep{luhman07}.  Therefore, SZ~Cha dominates the emission
in the GO spectra. In the GTO observations of ISO~52 a faint object is
present in the LL slit 8.5$^\prime$$^\prime$ from the target.  Due to the different orientation of the IRS
slit positions in the GO observations, this object is not present in the
LL slit in this epoch. However, this object is much fainter than ISO~52,
with a magnitude of about 14 in the 3.6 and 4.5~{\micron} IRAC bands (K.
Luhman, 2010, private communication), and so we can conclude that ISO~52
dominates the emission seen in its GTO spectrum.

\section{Analysis} \label{sec:ana}

\subsection{Flux Variability} \label{sec:var}

It is evident from the IRS spectra of our pre-transitional
(Figures~\ref{figirsptd1} and~\ref{figirsptd2}) and transitional objects
(Figure~\ref{figirstd}) that there is some variability in their fluxes. 
Here we quantitatively discern if there is true variability in the
sample and then qualitatively discuss the overall behavior of the
variability that is present.

\begin{figure}
\epsscale{1}
\plotone{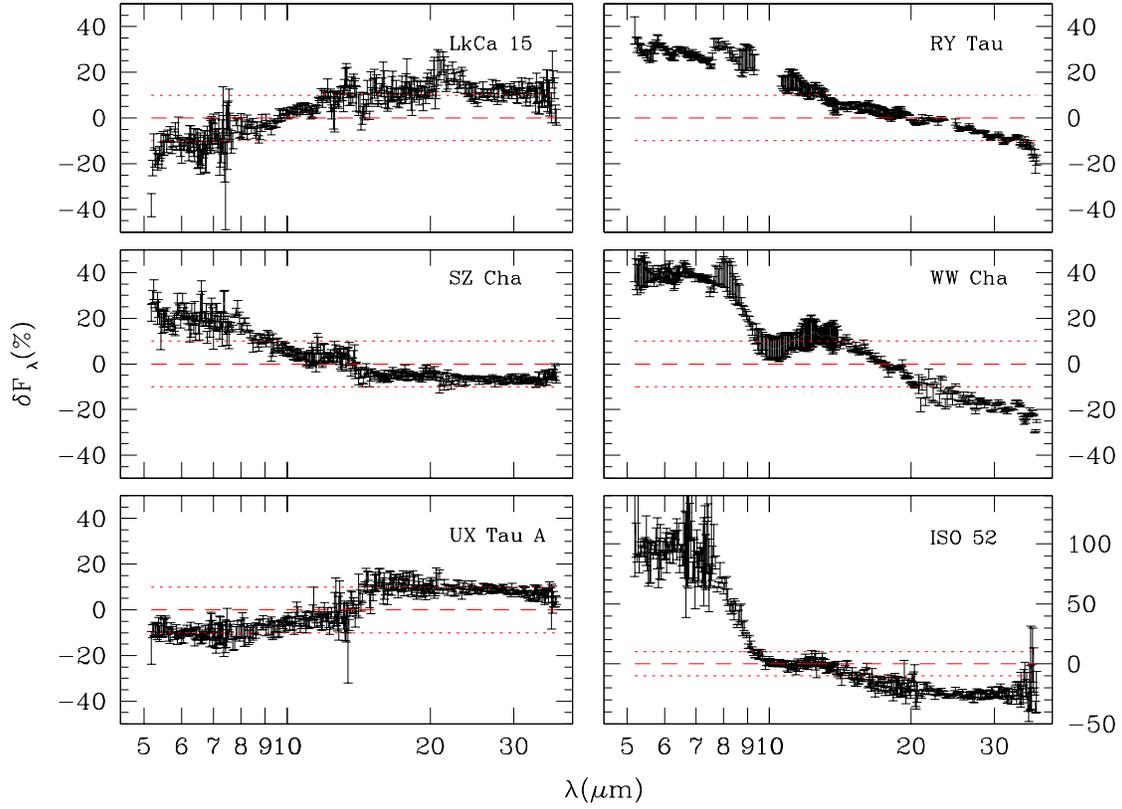}
\vskip -1.5in
\caption[]{Variability observed in the objects shown in Figure~\ref{figirsptd1}.  Here we show the percentage change in flux
between our GTO and GO1 spectra (i.e.
100~$\times$~(F$_{\lambda,GO1}$-F$_{\lambda,GTO}$)/F$_{\lambda,GTO}$). 
The error bars correspond to the uncertainties in the observations.  The dashed
line corresponds to no change in flux (i.e. $\delta$F$_{\lambda}\sim$0) and the
dotted lines correspond to a 10$\%$ change in flux.
In the case of UX~Tau~A,
we show the percentage change in flux between its GO1 and GO2 spectra (i.e. 
100~$\times$~(F$_{\lambda,GO2}$-F$_{\lambda,GO1}$)/F$_{\lambda,GO1}$). 
Note that the region around 10~{\micron} has been removed
in RY~Tau since its spectra were saturated in
this region.  The GTO observation of RY~Tau and the GO observations of WW~Cha were mispointed 
and so the variability seen here should be taken as an indication of the overall behavior 
of the variability in these objects and not as an accurate measurement of the amount of variability.
}
\label{figptd1err}
\end{figure}

\begin{figure}
\epsscale{1}
\plotone{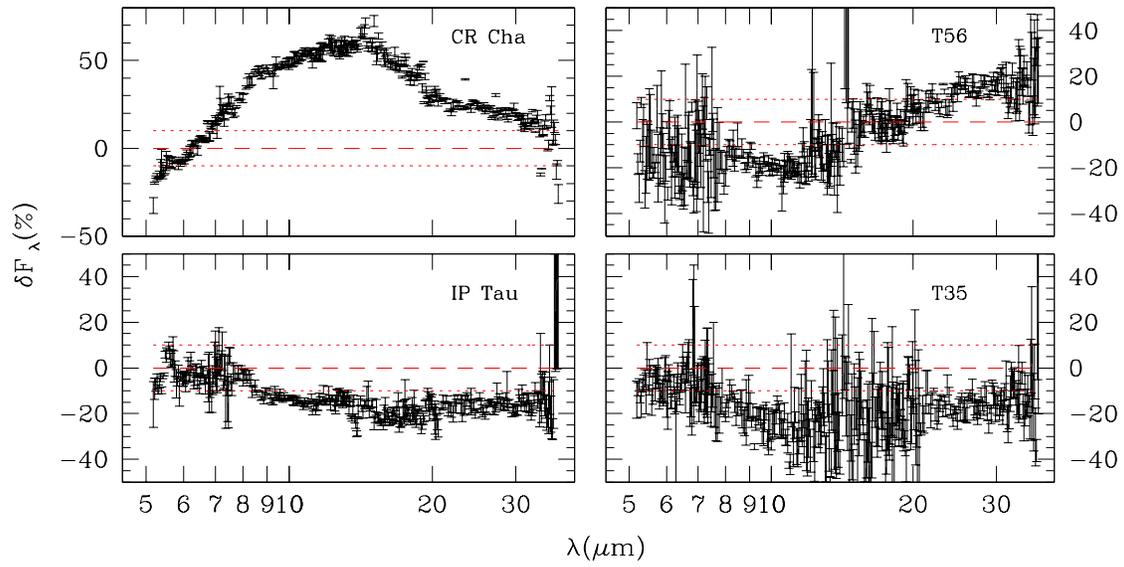}
\vskip -2.5in
\caption[]{Variability observed in the objects shown in Figure~\ref{figirsptd2}.   Symbols are the 
same as in Figure~\ref{figptd1err}.  In the case of T35,
we show the percentage change in flux between its GO1 and GO2 spectra.
}
\label{figptd2err}
\end{figure}

\begin{figure}
\epsscale{1}
\plotone{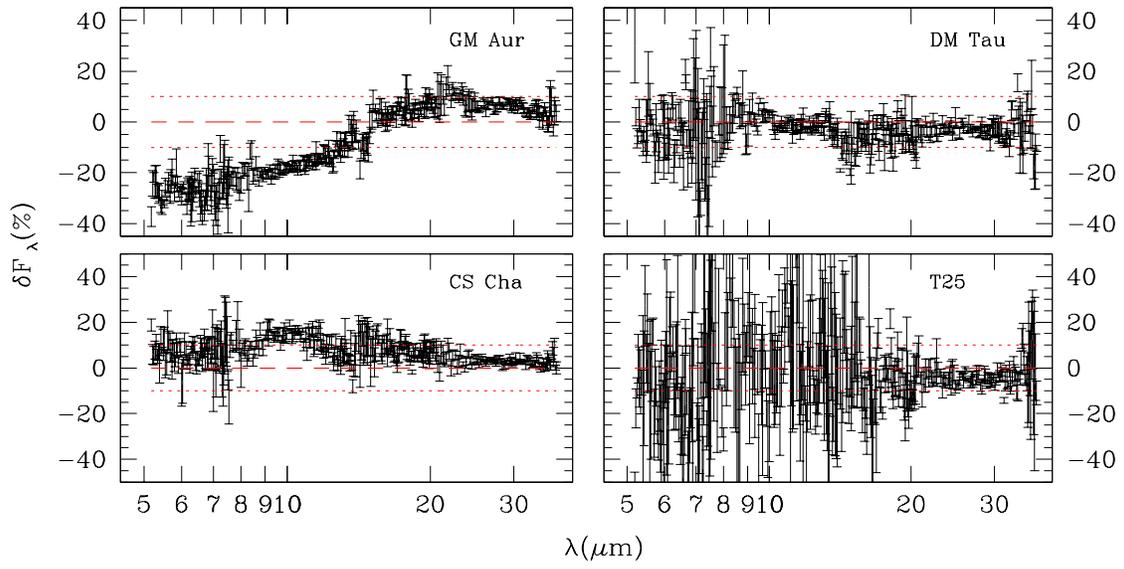}
\vskip -2.5in
\caption[]{Variability observed in the objects shown in Figure~\ref{figirstd}.  Symbols are the same as in 
Figure~\ref{figptd1err}.  Except in the cases of DM~Tau and T25, a $>$10$\%$ change in flux is 
observed.  
}
\label{figtderr}
\end{figure}

\begin{figure}
\epsscale{1}
\plotone{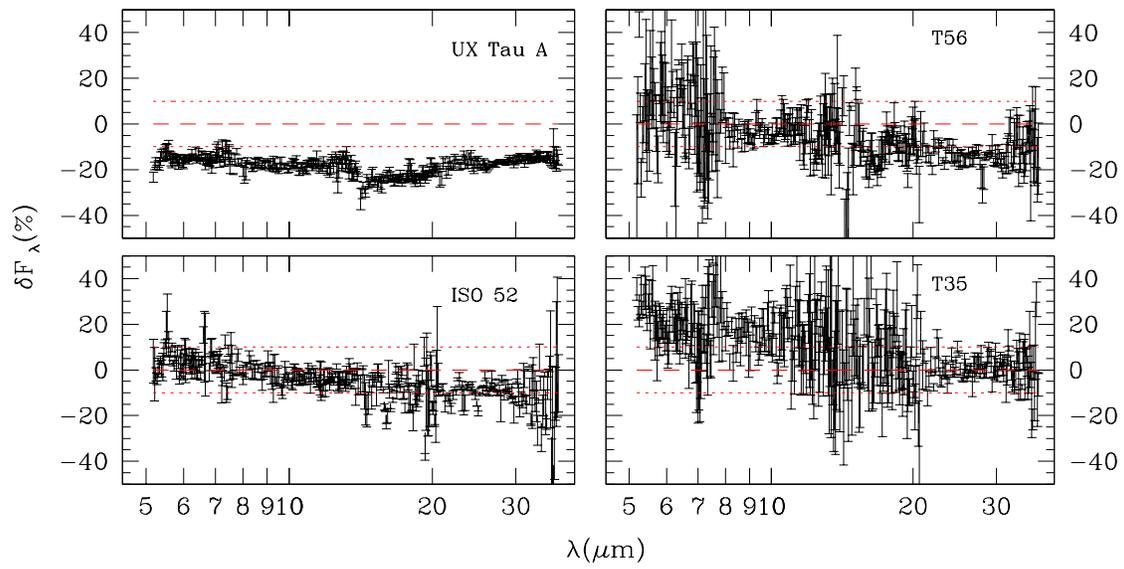}
\vskip -1.5in
\caption[]{Additional variability observed in UX~Tau~A, ISO~52, T56, and T35.  Here 
we show the percentage change in flux between the GTO and GO1 spectra for UX~Tau~A and T35.  For ISO~52 and T56 we plot
the difference between the GO1 and GO2 spectra.
}
\label{fig1wkerr}
\end{figure}

\begin{figure}
\epsscale{1}
\plotone{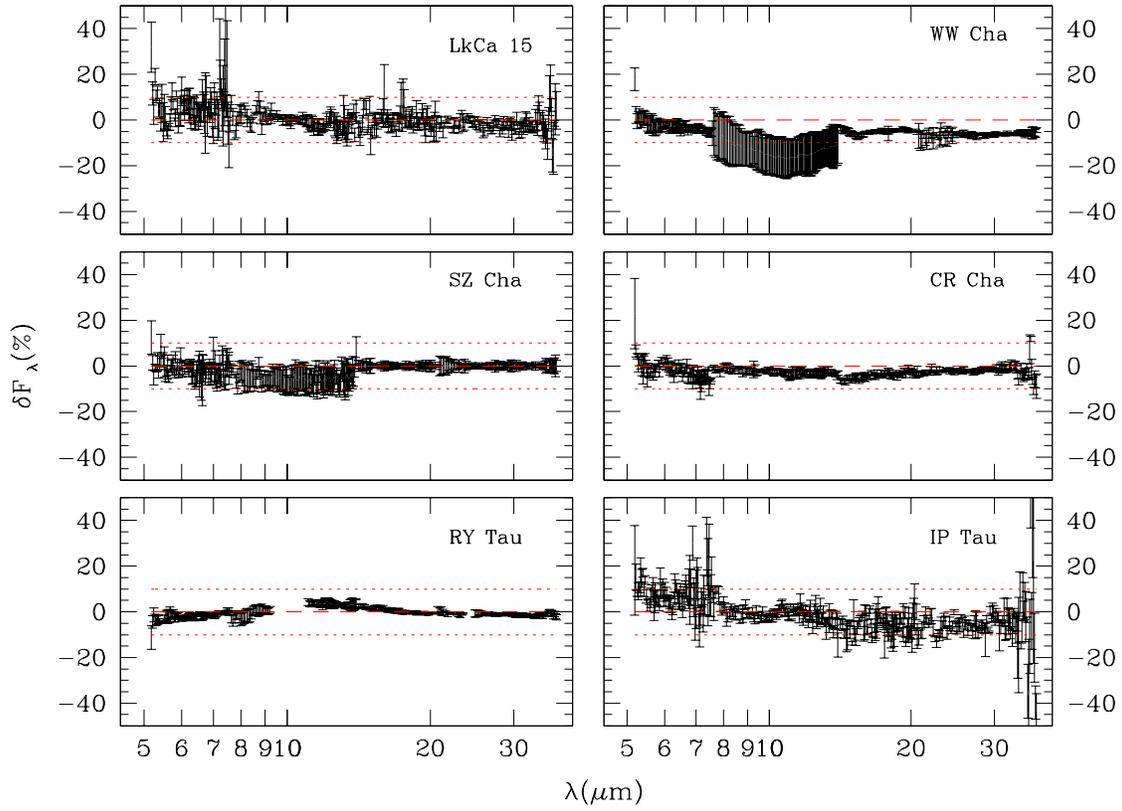}
\vskip -1.5in
\caption[]{Percentage change in flux between the GO1 and GO2 spectra for some of the pre-transitional disks in the sample.
None of these objects vary on this 1 week timescale.  We note that the WW Cha observations were significantly mispointed.  The percentage change in flux between the GO1 and GO2 observations of UX Tau A, ISO~52,
T35, and T56 can be found in Figures~\ref{figptd1err},~\ref{fig1wkerr},~\ref{figptd2err}, and~\ref{fig1wkerr}, respectively.
}
\label{fignovar1}
\end{figure}

\begin{figure}
\epsscale{1}
\plotone{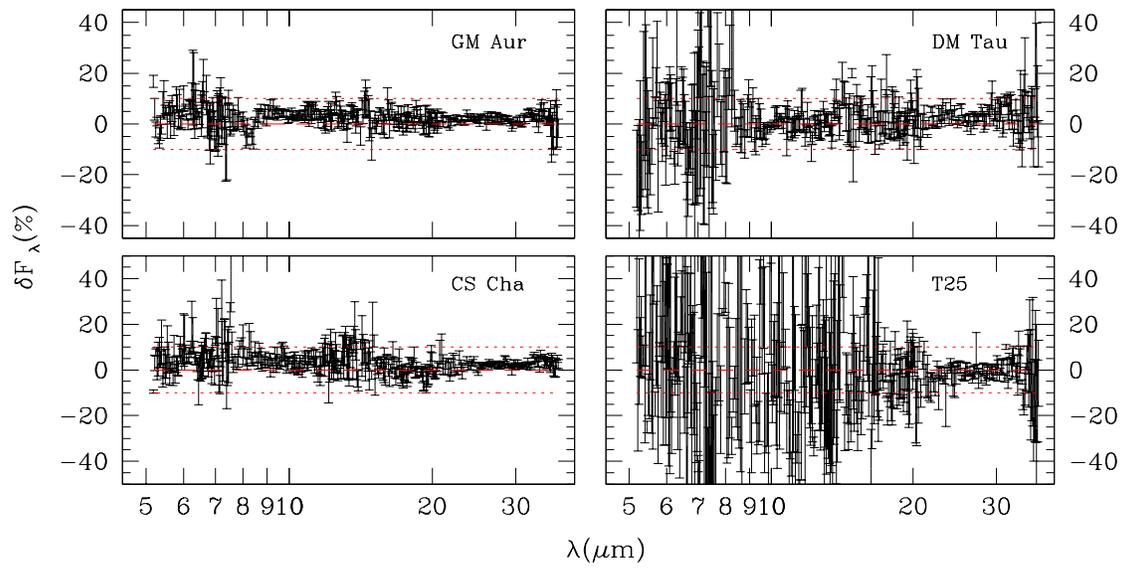}
\vskip -1.5in
\caption[]{Percentage change in flux between the GO1 and GO2 spectra for the transitional disks in the sample.
None of the objects vary on this 1 week timescale.
}
\label{fignovar2}
\end{figure}

Figures~\ref{figptd1err}--\ref{fignovar2} illustrate the change in
flux seen in an object in our sample between different epochs.
The differences in flux between the GTO and GO1 spectra are
shown in Figures~\ref{figptd1err}--\ref{figtderr}.  (Note: the same
analysis for UX~Tau~A and T35 is plotted in Figure~\ref{fig1wkerr}). 
The difference in flux ($\delta$F$_{\lambda}$) is plotted in terms of the
percentage difference in emission between the GTO and GO1 data relative
to the GTO data. The error bars in the figures correspond to the
uncertainties in the observations. Except for the cases of
DM~Tau and T25 (where $\delta$F$_{\lambda}\sim$0), we observe
significant variability outside of the observational uncertainties in
each of the targets between the GTO and GO1 observations, which were
taken more than a year apart. We performed the same analysis comparing
the GO1 and GO2 spectra, which were taken about one week apart, and we
only see variability on these timescales in four objects:  UX~Tau~A
(Figure~\ref{figptd1err}), ISO~52, T56 (Figure~\ref{fig1wkerr}), T35
(Figure~\ref{figptd1err}).  In Figures~\ref{fignovar1} and~\ref{fignovar2} we show that the 
rest of the objects in our sample do not vary significantly between the GO1 and GO2 epochs.

In many of the pre-transitional targets, the flux clearly oscillates
around a pivot wavelength (Figure~\ref{figptd1err}).  As the short
wavelength emission decreases, the emission at longer wavelengths
increases and as the short wavelength emission increases, the emission
at longer wavelengths decreases. The spectra of LkCa~15 can be taken as
representative of this group of objects.  In Figure~\ref{figptd1err} the
flux of its GO1 spectrum is $\sim$10$\%$ lower than the GTO spectrum at
wavelengths $<$8~{\micron}.  The fluxes of the two spectra are the same
around $\sim$10~{\micron}, but the GO1 spectrum is higher beyond that
point. In the other objects mentioned above, the overall behavior is
similar while the pivot wavelength and magnitude of the flux change can
vary from object to object. UX~Tau~A does not display this seesaw
behavior between the GTO and GO1 spectra (Figure~\ref{fig1wkerr}), but
it does between the GO1 and GO2 spectra (Figure~\ref{figptd1err}).  We
point out that while the RY~Tau GTO and WW~Cha GO observations were
mispointed, that should have resulted in an decrease of flux at all
wavelengths relative to the GO and GTO observations, respectively. 
However, we see flux losses at some wavelengths while flux gains at
others, indicating that there is true variability
in these objects, but we cannot accurately constrain the difference in
the flux or the pivot wavelength due to the mispointing.

Whether or not this seesaw behavior is present in the other
pre-transitional objects in the sample is unclear
(Figure~\ref{figptd2err}). The GO1 spectrum of CR~Cha has less emission
than the GTO spectrum shortwards of $\sim$6~{\micron}, but substantially
more at longer wavelengths.  Due to the artifacts in the GTO spectrum
$>$35~{\micron} discussed in {\S}~\ref{redux}, we cannot confidently
tell whether or not the GTO and GO spectra agree at these wavelengths.
In IP~Tau the flux at $\lambda$ $<$7{\micron} is about the same, but it
is lower for the GO1 spectrum at longer wavelengths. For T56, the
emission in the GO1 spectrum is higher than the GTO spectrum beyond
$\sim$20~{\micron}.  It appears that the flux in the GO1 spectrum is
lower at $<$20~{\micron}, but the SNR is too poor $<$7~{\micron} to tell
if this holds at the shortest wavelengths. In T35, when comparing the
GO1 and GO2 spectra, the flux is the same $<$7~{\micron}, but the GO2
spectrum has less emission beyond that.  Again, the spread in
uncertainties is large in T35 because of the poorer SNR.

The behavior observed in the transitional disks is displayed in
Figure~\ref{figtderr}. GM~Aur has the seesaw behavior seen in LkCa~15,
where the pivot is at $\lambda$$\sim$18~{\micron}. In CS~Cha, only the
flux of the 10~{\micron} silicate emission feature changes substantially
between the GTO and GO1 spectra. DM~Tau and T25 have no discernible
variability.

The four objects in our sample which vary on 1~wk timescales display behavior
that could be classified as seesaw-like as already described above, but
they also exhibit additional behavior (Figure~\ref{fig1wkerr}).
UX~Tau~A's GO1 spectrum is weaker at all wavelengths relative to the GTO
spectrum, indicating that the emission of this object has decreased with
time. The spread in uncertainties is large in ISO~52, T56, and T35 because of
poor SNR; however, it appears that the ISO~52 and T56 spectra diverge beyond
$\sim$20~{\micron} between the GO1 and GO2 spectra and that the T35
spectra diverge shortwards of $\sim$10~{\micron} between the GTO and GO1
spectra.

\subsection{Disk Model} \label{sec:mod}

Using disk models, we attempt to reproduce the SED variability observed
in Figures~\ref{figirsptd1}--\ref{figirstd}. The models used here are
those of \citet{dalessio98,dalessio99,dalessio01,dalessio05,dalessio06}. We refer the reader to
those papers for details of the model and to \citet{espaillat10} for a
summary of how we fit the SEDs of pre-transitional and transitional
disks in particular.

A full disk model has an irradiated accretion disk with a 
sharp transition at the dust sublimation radius.
We model this transition
as a frontally illuminated wall which
dominates the near-IR
emission. Pre-transitional disks have a gap in the
disk.  They have an inner disk separated from an outer disk by the gap.
The optically thick inner disk also has a sharp transition at the dust sublimation
radius, as seen in full disks, which we model as an inner wall.  
 In the subsequent modeling analysis, we do not include the
contribution to the SED from the inner disk behind this wall since
previous work has shown that the inner wall dominates the emission at
these shorter wavelengths. There is another wall located where the outer
disk is inwardly truncated (i.e. the outer edge of the gap) and this
outer wall dominates the SED emission from $\sim$20--30~{\micron}.
Behind this wall, there is an outer disk which dominates the emission
beyond $\sim$40~{\micron}. Since transitional disks have holes in their
disks, they do not have the inner wall seen in pre-transitional disks.
When modeling transitional disks, we include the outer wall and outer
disk described above. In both pre-transitional and transitional disks,
the gap or hole sometimes contains a small amount of optically thin dust
which dominates the contribution to the 10~{\micron} silicate emission
feature. We calculate the emission from this optically thin dust region
following \citet{calvet02}. We note that in the case of the
pre-transitional disks, the inner optically thick disk will cast a
shadow on the outer disk and here we include the effect of this
shadowing on the outer wall following \citet{espaillat10}.
In short, since the star is a finite source, there is both a penumbra and umbra on the outer wall.
In the umbra, the wall is not illuminated and in the penumbra, the wall is partially illuminated.  Above
the penumbra, the wall is fully illuminated. 
Refer to the Appendix of \citet{espaillat10} for more details.

\subsubsection{Stellar Properties}

Table~\ref{tab:stellar} lists stellar properties for our sample which
are relevant for the disk model. 
We note that the stellar properties of our objects are based on optical and near-infrared
data which are not contemporaneous with the IRS spectra analyzed in this work.  
If the star's properties change over time this can
result in uncertainties in the input stellar parameters and hence the disk properties derived here.

Spectral types for our objects are from
the literature and the temperature for the spectral type listed in
Table~\ref{tab:stellar} was taken from \citet{kh95}. The stellar
properties for our sample (L$_{*}$, M$_{*}$, R$_{*}$) are from the HR
diagram and \citet{siess00} tracks. When U-band photometry was
available, the mass accretion rates were derived in this work using
U-band data and the relation in \citet{gullbring98}. Extinction
corrections were made by matching V- , R- , I-band, and 2MASS photometry
to photospheric colors from \citet{kh95}.  The spectra were dereddened
with the \citet{mathis90} dereddenning law. The distance adopted for
Taurus is 140~pc \citep{bertout99} and for Chamaeleon this is 160~pc
\citep{whittet97}.

\subsubsection{Disk Properties}

Table~\ref{tab:disk} lists the disk properties of our sample.  When
parameters are specific to only one epoch, this distinction is made in
the table (see table footnote). We assume that the inclination of the
disk is 60 degrees, unless a measurement could be found in the
literature. $T_{wall}$ is the temperature at the surface of the
optically thin wall atmosphere. The temperature of the inner wall
(T$_{wall}^i$) is typically held fixed at 1400~K (except in the cases of
UX~Tau~A and T35 which will be addressed in the Appendix). The
temperature of the outer wall (T$_{wall}^o$) is varied to fit the SED
best.  The radius of the wall ($R_{wall}$) is derived using $T_{wall}$
following Equation 2 in \citet{espaillat10}. The heights of the walls
(z$_{wall}$) and the maximum grain sizes (a$_{max}$) are adjusted to fit
the SED. The parameters of the outer disk are also varied to fit the
SED. These include the viscosity parameter ($\alpha$) and the settling
parameter ($\epsilon$; i.e. the dust-to-gas mass ratio in the upper disk
layers relative to the standard dust-to-gas mass ratio).  M$_{disk}$ is
calculated according to Equation 38 in \citet{dalessio98} and it is
proportional to ${\mdot}/{\alpha}$. We adopt an outer disk radius of
300~AU for all of our disks.

\subsubsection{Dust Opacities} \label{sec:moddustopa}

As discussed in \citet{espaillat10}, the opacity of the disk has
important consequences on the resulting SED.  The opacity is affected by the
sizes of the dust grains and the composition of the
dust used. The grain size distribution used in the models follows the
form $a^{-3.5}$ where $a$ varies between $a_{min}$ and $a_{max}$
\citep{mathis77}.  
We assume the grains are spherical and note that while
irregularly shaped grains may have different opacities from spherical grains \citep{min07}, it is outside the scope of this work to constrain the shape of the dust
grains. 
Throughout the disk,  $a_{min}$ is held fixed at
0.005~{\micron}. In the walls, $a_{max}$ is varied to achieve the best
fit to the SED. 
We try maximum grain sizes between 0.25~{\micron} and 10~{\micron}.  The
wall emission is primarily optically thick, but also has an optically
thin component from the wall atmosphere which contributes to the
silicate emission features. Smaller grain sizes lead to a strong, narrow
10~{\micron} silicate emission feature while larger grain sizes produce
wider and less prominent emission features \citep[see Figure 3 in][]{espaillat07a}. In the outer disk, there are two dust grain size
distributions in order to simulate dust growth and settling 
\citep[see][for more details]{dalessio06}. In the upper disk layers,
$a_{max}$=0.25~{\micron} and in the disk midplane the maximum grain size
is 1~mm \citep{dalessio06}. For the dust composition of the inner wall
we follow \citet{dalessio05} and \citet{espaillat10} in adopting
silicates with a dust-to-gas mass ratio (${\zeta}_{sil}$) of 0.0034.   We note that only silicates
exist at the high temperatures at which the inner wall is located.
There are other types of dust such
as metallic iron that can exist at high temperatures  \citep{pollack94}.  
However, here we adopt a dust composition consistent with the one proposed by \citet{pollack94} 
for accretion disks.

We perform a more detailed dust composition fit for the silicates in the
outer wall and disk than done in our previous works.  The motivation is
that for a variability study trying to trace small changes in the flux,
it is important to isolate the continuum emission of the disk.  By
fitting the silicate dust features seen in the IRS spectrum as closely
as possible, one can then more clearly see the effect of changing the
disk continuum. Here we adopt a dust-to-gas mass ratio (${\zeta}_{sil}$)
of 0.0034 for the silicates in the outer wall and disk and explore
silicate dust mixtures incorporating olivines, pyroxenes, forsterite,
enstatite, and silica. (We note that throughout this work we are
referring to amorphous material of olivine or pyroxene stoichiometry
when using the terms ``olivine'' and ``pyroxene.'') We list the derived mass
fractions in Table~\ref{tab:silwall}. 

The optical constants used for
olivines and pyroxenes come from \citet{dorschner95}.  We calculate the
opacities assuming segregated spheres and Mie theory for the adopted dust grain
size distribution.
For an explanation of how the forsterite opacity was computed, see the discussion
by \citet{poteet10}.

We also calculated the opacity for enstatite.  We adopt optical constants for
enstatite from \citet{huffman71} and  \citet{egan77},
crystalline bronzite at 300K from  \citet{henning97}, the three
crystalline axes of orthoenstatite from  \citet{jaeger98}, and crystalline
hypersthene from \citet[][Sample 1S]{jaeger94} for the 0.1--0.5, 0.533--1.105,
6.7--8.4, 8.7--98, and 98--8000~{\micron} wavelength regimes, respectively.  The
optical constants from \citet{jaeger94} were modified to match the values
from \citet{jaeger98} as was done for forsterite as described by \citet{poteet10}.  Beyond 585~{\microns}, 
the real part of the index of refraction, $n$, was chosen to be a constant value equal to
the value of $n$ at 585~{\microns} obtained by modifying the \citet{jaeger94} $n$
values, and the 
imaginary part of the index of refraction, $k$, was determined by scaling a 1$/{\lambda}$ curve to the value of $k$, at
585~{\microns} from the modified \citet{jaeger94} values.  The absorption
opacity was then computed from these optical constants by CDE theory \citep{bohren83}.
The scattering opacity is assumed to be zero.

Finally, we compute the opacity for silica.  We adopt optical constants from the
following sources for silica.  Between 0.05 to 0.15~{\microns}, alpha quartz from
\citet{palik85} is used.  From 3-8~{\microns}, $k$ comes from $k_{abs}$ for amorphous
silica from \citet{palik85}.  For $k$ between 0.15 and 3~{\microns}, $k$ is interpolated
between its values at 0.15 and 3~{\microns}.  From 0.15 to 5.5~{\microns}, $n$ comes
from \citet{palik85} for alpha quartz.  Between 8 and 30~{\microns}, the $n$ and $k$
values for beta quartz at 975~K from \citet{gervais75} are used.  From
50-333~{\microns}, both $n$ and $k$ are from \citet{loewenstein73} for alpha quartz
at room temperature.  The value of $n$ at 333~{\microns} was kept constant to 8000~{\microns}.  For $k$, a 1$/{\lambda}$ curve fit to the value of $k$ at 333~{\microns} was used.
To compute the absorption opacity, we employed CDE at all wavelengths except 8 to 40~{\microns}. Between
8 to 40~{\microns} the absorption opacity was for annealed silica from \citet{fabian01}.  
The scattering opacity at all wavelengths is assumed to be zero.

In addition to silicates, for
each of the disks, we add organics and troilite to the dust mixture
following \citet{espaillat10} with ${\zeta}_{org}$ = 0.001 and
${\zeta}_{troi}$ = 0.000768 and sublimation temperatures of $T_{org}$ =
425~K and $T_{troi}$ = 680~K. We include water ice as well with a
sublimation temperature of 110~K.  Unless otherwise noted, we use
${\zeta}_{ice}$ = 0.00056. Optical constants for organics, troilite, and water
ice are adopted from \citet{pollack94}, \citet{begemann94}, and
\citet{warren84}. In objects where we include optically thin dust within
the hole or gap, the silicate dust composition is listed in
Table~\ref{tab:silthin}.  The abundances of silicates, organics, and
troilite in the optically thin dust region are given in the subsequent
sections.  We do not include ice in the optically thin region since the
temperatures here are high enough for it to have sublimated.

\subsection{SED Modeling} \label{sec:modsed}

Here we provide an overview of our modeling results.
In the Appendix, we describe in detail the modeling conducted in this
study for each individual object.  

\subsubsection{Disk Structure}

We can explain most of the seesaw variability observed in the pre-transitional disks
by changing the height of the inner disk wall. (We note that other possible explanations
including changes in the stellar and disk properties have not been explored here.)
In the pre-transitional disks of LkCa~15,
SZ~Cha, and UX~Tau~A we can reproduce the
seesaw variability by changing the height of the inner
disk wall by $\sim$22$\%$, 33$\%$, and 17$\%$ respectively (Figure~\ref{figlkca15}).
When the inner wall is taller, the emission at the shorter wavelengths where the wall
dominates the emission is higher; there is also a larger shadow on the outer wall and hence the
emission seen from the outer wall is less and the IRS spectrum is lower.   Correspondingly, when the inner wall is lower there is less near-IR emission and the shadow on the outer wall is smaller and so we see more emission from the outer wall longwards of 20~{\micron}.  The
10~{\micron} silicate emission in LkCa~15 and SZ~Cha does not change. This emission is
dominated by small dust in the optically thin region.  UX~Tau~A does not have
a discernible 10~{\micron} silicate emission feature.

Because of uncertainties
introduced to the observations by mispointing, we do not attempt to reproduce the
variability seen in RY~Tau (Figure~\ref{figrytau}). 
However, we can fit the SED of RY Tau with an 18~AU gap which contains some
optically thin dust.  In the case of WW~Cha (Figure~\ref{figwwcha}), due to the fact that the GO
observations were significantly mispointed and that we do not have a mass accretion rate estimate, we do not attempt to model its disk here.

We also do not have a mass accretion rate for ISO~52 (Figure~\ref{figiso52}).  However, these observations
were well pointed and for the purposes of reproducing
the general trend seen in the variability, we assume a typical value (see the Appendix for more details).
In this object, we need to increase the height
of the inner wall by $\sim$400$\%$ between the GTO and GO epochs 
to explain the observed variability.   Assuming that
our assertion that the inner wall height is varying is correct, this is by far the largest
change in wall height seen in the sample.

CR~Cha has a substantial change in slope at $\sim$6~{\micron} (Figure~\ref{figchx3}), from which
one could
infer that there is either a substantial change in the temperature of the wall or a change in the nature of the emission from
optically thick to optically thin.  We find that
CR~Cha is best explained with a pre-transitional disk model in the GTO
observations and either a pre-transitional or transitional disk model in the GO observations.
See the Appendix for more details and {\S}~\ref{crchadis} for a discussion.

In the pre-transitional disks of IP~Tau, T56, and T35, we can reasonably reproduce the
emission within the uncertainties of the observations by varying the height of the inner
disk wall by 17$\%$, 50$\%$, 
and 20$\%$, respectively (Figure~\ref{figiptau}). 
In addition to varying the height of the inner
wall, we also had to change the amount of dust in the optically thin
regions of IP~Tau and T56 in order to reproduce the variability in the
10~{\micron} silicate emission feature.

We also modeled the transitional disks in the sample (Figures~\ref{figgmaur}
and~\ref{figdmtau}).  GM~Aur displays seesaw behavior. To fit it we vary the amount of
optically thin dust in the hole and have to change the height of the
outer wall as well.  In CS~Cha, only
the 10~{\micron} silicate emission changes between epochs and to fit 
this variability we alter the amount of dust in the optically thin
region. In DM~Tau and T25 there is no variability and there
is no evidence for significant amounts of dust in their holes.

There is some variability in the pre-transitional disks which we cannot 
explain by changing the height of the inner disk wall.
The GO2 spectrum of T35 has a change in slope at $\sim$7~{\micron} which we
could not explain with the disk models presented here.  We have no
obvious explanation for this but speculate it could be related to the
high temperature derived for the inner disk wall (1800~K).
In UX~Tau~A, the overall emission from the disk 
has decreased with time (i.e. the GO spectra have less emission than the
GTO spectra).  
While we do not try to fit this decrease with models,  it
can possibly be attributed to a decrease in the luminosity  at the bands where the disk absorbs
stellar radiation, most likely due to star spots \citep{skrutskie96}, or
an overall decrease in the accretion luminosity of the disk, most likely from a change in
the mass accretion rate by a factor of about 3. 
We also analyzed an additional SHLH spectrum from the {\it Spitzer} archive (see Appendix for more details).  This spectrum has substantially lower emission at ${\lambda}>$13~{\micron} (Figure~\ref{uxshlh}).  
We can fit the SHLH spectrum using an inner wall
with a temperature of 1800~K.  This hotter inner wall is closer to the star and leads to a larger shadow on the outer wall.  
Given that we do not have simultaneous data at shorter wavelengths, we  
cannot test if this wall fits the SED at ${\lambda}<$10~{\micron} at the time the SHLH spectrum was taken.

\subsubsection{Dust Composition}

As a result of trying to reproduce the variability observed in our
sample, in this study we also performed fitting of the silicate
emission features visible in the IRS spectra, deriving the mass fraction
of amorphous and crystalline silicates in the outer wall and the
optically thin regions (Tables~\ref{tab:silwall} and~\ref{tab:silthin}).
(We refer the reader to Figure 1 of \citet{watson09} for the positions
of the strongest features of crystalline silicates visible in IRS spectra.)
We do not attempt a detailed ${\chi}^{2}$ fitting since it would be too computationally
expensive to do so with our disk code.  Thus, the derived mass fractions in
Tables~\ref{tab:silwall} and~\ref{tab:silthin} should be taken as
representative of a dust composition that can reasonably explain the
observed SED. 
We refer the reader to \citet{sargent09} for
a review of the typical degeneracies of dust fitting.  In short, large grains of amorphous olivine and amorphous pyroxene composition are the most degenerate, in the sense that one of these components
could be replaced by the other and a similar fit would be found.  Enstatite and forsterite are
also somewhat degenerate at cooler temperatures.  Another caveat, which was noted earlier,
is that the shapes of the grains in the disk are not well known.
We leave it to future work to further constrain the mass fractions of silicates in these disks.

For the inner wall of our objects, the silicate composition consisted
solely of amorphous olivines.  However, since the inner wall does not
produce significant 10~{\micron} silicate emission in the objects in
this study, we have no way to distinguish between pyroxene and olivine
silicates in the inner wall.  Also, while we included crystalline
silicates in the disk behind the outer wall, it is the outer wall that
dominates the emission at the longer IRS wavelengths.  Because of these
previous two points, here we only discuss the composition of silicates
in the optically thin dust region and the outer wall.

Most of the absorption and emission of the outer walls in our sample are dominated by amorphous
silicates (Table~\ref{tab:silwall}).  The exception is T35 which is dominated by crystalline
silicates ($\sim$60$\%$). The optically thin region also tends to be dominated by
amorphous silicates with typically $\sim$10$\%$ or less of crystalline
silicates.  This is not the case in T56 which contains $\sim$25$\%$
crystalline silicates. Of the three crystalline silicates studied in
this work, we are more likely to see forsterite rather than
enstatite and silica in the optically thin region (Table~\ref{tab:silthin}). Comparing the
optically thin region and outer wall in objects that have both, it
appears that silica is more likely to be present in the outer wall.
Relative to the optically thin region, we find more crystalline
silicates in the outer walls of T56, SZ~Cha, and LkCa~15 and less for
CS~Cha and GM~Aur. The amount of crystalline silicates in CR~Cha and
IP~Tau is the same between both regions. RY~Tau has no evidence for
significant amounts of crystalline
silicates in its disk.

The results from the dust fitting performed in this work are in
reasonably good agreement with the detailed dust fitting conducted by
\citet{sargent09} which used a two-temperature model.  The objects that
the two samples have in common are DM~Tau, GM~Aur, IP~Tau, and LkCa~15. 
 Both works find that these four disks are dominated by amorphous
silicates and that there are relatively few crystalline silicates
present.  Furthermore, olivine silicates dominate the inner parts of the
disk that contribute to the 10~{\micron} emission.

\section{Discussion} \label{sec:discuss}

\subsection{Linking Infrared Variability to Disk Structure}

Understanding the underlying causes of the variability observed in this
sample depends upon the physical locations in the disk from which the
changes in flux arise.  Given that the sample was chosen to include
pre-transitional and transitional disks, the nature of these objects
will necessarily play a key role in this. The disk structures of
LkCa~15, UX~Tau~A, GM~Aur, and RY~Tau have been independently confirmed.
LkCa~15, UX~Tau~A, and GM~Aur have been imaged with millimeter
interferometers and large cavities in their disks have been observed
\citep[Andrews et al, in prep;][]{pietu06,hughes09}. Near-infrared
spectra have confirmed that the inner disks of LkCa~15 and UX~Tau~A are
optically thick while the inner disk of GM~Aur is optically thin
\citep{espaillat10}. Millimeter interferometric imaging of RY~Tau by
\citet{isella10} detects two spatially resolved peaks, an indicator of a
disk hole, whose separation translates to a cavity that is consistent
with the 18~AU gap inferred from the SED modeling in this work. For the
other objects in the sample, the disk structure is inferred solely from
SED modeling. Millimeter interferometry and near-IR data are needed to
confirm that there are cavities in these disks and to probe if the inner
disk is optically thick. However,  the IRS spectra of SZ~Cha, WW~Cha,
and T56 are reminiscent of LkCa~15, suggesting that they are gapped
disks as well.  Likewise, T35 resembles UX~Tau~A.  CS~Cha, DM~Tau, and
T25 have large deficits of flux which are strong indicators of inner
disk holes, as seen in GM~Aur.

It follows that one can roughly divide the disk into two regions --
inner (inner wall and$/$or optically thin dust region) and outer (outer
wall and outer disk). Interestingly, the only disks that do not display
variability in our sample are the transitional disks DM~Tau and T25
whose inner regions do not contain substantial amounts of small dust. 
DM~Tau's inner hole is relatively devoid of small
dust and T25's inner region  contains only 10$^{-13}$~{\msun}.  In
contrast, the transitional disks of GM~Aur and CS~Cha have about ten times more small dust within
their optically thin inner cavities than the transitional disks of DM~Tau and T25. We can infer that there is not
enough material in the inner regions of DM~Tau and T25 to lead to
significant variability.  Objects in the sample that have a notable
amount of material in their inner region do vary.

\subsubsection{Inner Wall}

We attempted to understand the variability seen in our sample by fitting the
SED with disk models.
In the pre-transitional disks
of LkCa~15, SZ~Cha, UX Tau~A, IP~Tau, T56, and T35 we can reasonably
reproduce the emission between 5--38~{\micron} within the uncertainties by varying the height of
the inner disk wall by 22$\%$, 33$\%$, 17$\%$, 17$\%$, 50$\%$, 
and 20$\%$, respectively (Figures~\ref{figlkca15} and~\ref{figiptau}). When
the inner wall is taller, the emission at the shorter wavelengths is
higher since the inner wall dominates the emission at 2--8~{\micron}.
The taller inner wall casts a larger shadow on the outer disk wall and
we see less emission at the wavelengths beyond 20~{\micron} where the
outer wall dominates.  When the inner wall is shorter, the reverse
occurs. ISO~52 is an extreme case.  Its inner wall height has to change by
400$\%$ to explain the observed variability (Figure~\ref{figiso52}).

We did not attempt to fit the variability seen in the pre-transitional
disks of RY~Tau and WW~Cha due to complications introduced by
mispointing and insufficient data. However, these disks exhibit
seesaw-like variability
(Figures~\ref{figrytau} and~\ref{figwwcha}). Taking the
modeling described above into consideration, one can surmise that the
variability in these disks is also due to an inner wall which varies in
height.
While the variations in the SED for many objects in the sample can be reproduced by changes in the
height of the inner wall, we note that other explanations that have not been considered here
may possibly result in similar SED behaviors.  We leave exploration of this to future work.

\subsubsection{Optically Thin Dust Region}  \label{disoptthin}

In the pre-transitional disks of IP~Tau and T56 the 10~{\micron}
silicate emission changes (Figure~\ref{figiptau}).  This feature is
dominated by sub-micron sized grains in the optically thin dust region
located within the disk gap.  We can reproduce the change in this
emission by adjusting the amount of small dust in this region.
Alternatively, given that the spatial distribution of this dust is
largely unknown, it is possible that part of the optically thin dust
region is in the shadow of the inner wall and so in some cases the
amount of dust we see in this region varies as the height of the inner
wall changes. In LkCa~15 and SZ~Cha, the 10~{\micron} silicate emission
does not change. This indicates that the optically thin dust is
vertically distributed in such a way that it is not shadowed by the inner wall. 
This could suggest that there is more dust in the gap that we do not
detect. Therefore, the values for the amount of dust in the gaps of
pre-transitional disks should be taken as a lower limit.  Alternatively,
this optically thin dust could be heated indirectly.
In order to explain the presence of the 10~{\micron} silicate emission
feature in self-shadowed Herbig Ae$/$Be stars,
\citet{dullemond04} proposed that light reaches the shadowed 
regions after being scattered off of the upper parts of the inner wall.  
In addition, they suggest that the thermal emission from the wall 
may also irradiate the shadowed region.  Similar mechanisms may
be at work in the optically thin regions of pre-transitional disks.

Of the four transitional disks in the sample, two exhibit variability:
CS~Cha and GM~Aur (Figure~\ref{figgmaur}). In CS~Cha, only the
10~{\micron} silicate emission changes between epochs. We can explain
this by varying the amount of optically thin dust located within the
central cavity.  On the other hand, the dust in CS~Cha could be
spatially distributed in such a way as to cause the observed
variability. For example, CS~Cha is a spectroscopic binary and, while we
do not see the type of variability expected for a circumbinary disk
\citep{nagel10},  the optically thin dust in the hole of CS~Cha could be
unevenly distributed in such a way that the alignment of the binary
system at the time when the GO data was taken illuminates more of the
dust. High-resolution near-infrared interferometry of this object would
be ideal to test this.

To fit the variability of GM~Aur, we not only change the amount of
optically thin dust in the hole, we have to change the height of the
outer wall as well. \citet{espaillat10} demonstrated that GM~Aur's
near-infrared excess continuum between 1--5~{\micron} could be
reproduced by emission from sub-micron-sized optically thin dust.  These
variability data suggest that there is some optically thick structure
in the inner disk 
perhaps composed of large grains 
and$/$or limited in spatial extent 
which does not contribute
substantially to the emission between 1--5~{\micron} and leads to
shadowing of the outer disk.  Alternatively,  it could be that while the
dust in the hole is vertically optically thin, it becomes horizontally
optically thick at some radius and shadows the outer disk
\citep{mulders10}.  This scenario implies that the vertical extent of 
the optically thin dust region changes to produce a shadow commensurate with the variable emission
at the longer wavelengths.  
Then again, the fact that we have to change the height of the outer wall of GM Aur to fit its variability may not be linked
to shadowing of the outer wall by inner disk material, but by changes in the outer wall itself.  For example, our models assume the outer wall is axisymmetric, but changes in the visible area of the wall could lead
to different emission from the outer wall.  We note however that the orbital timescales at the outer wall are much
longer than the timescales probed in this work.
The case of GM~Aur needs to be explored further.

\subsubsection{The Unique Case of CR Cha} \label{crchadis}

CR~Cha displays a considerable change in slope at 6~{\micron}
(Figure~\ref{figchx3}). In the other objects in this sample, the
emission at this wavelength is typically dominated by either an
optically thick inner wall or an optically thin dust region. This could
suggest that the inner disk alternates from being dominated by
optically thick material in the GTO epoch to being dominated by
optically thin dust at the time that the GO observations were taken
3~yrs later.  Alternatively, the change in slope at 6~{\micron} could
also be due to a substantial decrease in the temperature of the inner
wall.
Accordingly, we reproduced the variability observed in
CR~Cha by fitting it with a pre-transitional disk model in the GTO epoch
and both a transitional and pre-transitional disk model in the GO epoch. 

In the case where we fit the GO spectra with a transitional disk model, it follows that once the
optically thick inner wall disappears, we see all of the optically thin
dust within the disk hole.  Hence there is substantially more
10~{\micron} emission in the GO observations which we reproduce by
increasing the amount of dust in the optically thin region.
We note that the height of the outer wall in the GO epoch, where we assume
that there is no shadowing, is less than the height of the wall in the GTO epoch
when there is an optically thick inner wall shadowing part of the outer wall.
Given that we take the shadowed portion of the outer wall into account in the case where the inner disk is optically
thick, this decrease in wall height in the case where the inner disk is optically thin
may imply that a portion of the outer wall is still shadowed.
As in the case of GM Aur discussed in {\S}~\ref{disoptthin}, this suggests
that either there is an optically thick structure in the inner disk that we cannot detect or that the vertically optically thin dust is radially optically thick.
It is expected that the optically thick inner disk in pre-transitional
disks will disappear at some point via accretion onto the star and$/$or
a lack of resupply of dust and gas from the outer disk, leaving behind a
transitional disk.   However, the viscous timescale at these radii
is on the order of 10$^{4}$~yrs making it improbable that we are detecting
this transition.  

Alternatively, in the case where we fit the GO spectra with a pre-transitional disk model, we have to 
decrease the temperature of the wall from 1400~K (in the GTO fit) to 800~K.  This corresponds
to a change in radius from 0.2~AU to 1~AU.  This may not be a viable model, since it is not clear
what process could make the dust grains sublimate above 800~K.  In any event, the amount of optically thin dust in the gap
remains about the same as seen in the GTO epoch.

Near-IR spectra at shorter wavelengths are necessary to decipher whether or not
the inner disk of CR~Cha is optically thick or optically thin.  Multi-epoch spectra would
be useful in constraining if the nature of the inner disk changes with time.
Millimeter confirmation of the hole in CR~Cha with
{\it ALMA} is necessary to decipher if there is indeed a cavity in this
disk.

\subsection{Physical Mechanisms Behind Variable Disk Structures}

\subsubsection{Variable Accretion}

A higher mass accretion rate will lead to a higher surface density in
the disk and so the height of the wall (defined as the point where the
optical depth to the stellar radiation reaches $\sim$1) will increase
\citep{muzerolle04}. In the cases of LkCa~15, SZ~Cha, UX Tau~A, IP~Tau,
T56, and T35  the change in the near-IR emission could be explained if
the mass accretion rate varies by factors of $\sim$3--10 relative to the
mass accretion rate used in this work. Studies have shown
that mass accretion rates onto the star are indeed variable. In the
transitional disk of TW Hya, accretion rates of 5$\times 10^{-10}$
M$_{\sun}$ yr$^{-1}$ (Muzerolle et al. 2000), 2$\times 10^{-9}$
M$_{\sun}$ yr$^{-1}$ (Herczeg et al. 2004), and 3.5$\times 10^{-9}$
M$_{\sun}$ yr$^{-1}$ (Ingleby \& Calvet, submitted) have been measured.
Alencar \& Batalha (2002) found that TW Hya's mass accretion rate varied
between 10$^{-9}$ -- 10$^{-8}$ M$_{\sun}$ yr$^{-1}$ over a one year
period and that smaller variations were seen even on periods of days.
While these accretion rates have been measured onto the star (i.e. the accretion columns),
the orbital timescales at the dust sublimation radius ($\sim$1~week) are within the timescales of
infrared variability seen. Hence the changes in the mass accretion rate in the inner disk necessary to change the
wall height are plausible.

However, if the mass accretion rate increases, the radius of the dust
sublimation radius will increase as well given that
$R_{wall}\propto(L_* + L_{acc})^{0.5}$ and $L_{acc}$ $\sim GM_*\mdot/R_*$ \citep{dalessio05}. 
The change in the radius is much larger than the change in the wall height.
For a wall with a relatively similar height and a larger radius, the
shadow on the outer wall \citep[see Equations A4 and A5 in the appendix of][]{espaillat10} will not
be large enough to diminish the flux at the longer wavelengths to
the levels observed.  Therefore, a change in mass accretion rate alone
cannot explain the observed SEDs in the pre-transitional disks in our
sample, indicating that the variability in these disks is due to a
change in the wall height while keeping the radius of the wall fixed.

Earlier we noted that to fit the variability in the transitional disks
of  GM~Aur and CS~Cha and the pre-transitional disks of IP~Tau and T56,
we altered the amount of dust in the optically thin region.  It has been
proposed that this small dust exists in the holes of some objects due to
dust traveling with gas from the outer disk and into the inner disk
after being filtered at the outer wall \citep{rice06}.    In this
scenario, changes in the amount of dust in the optically thin region
could be due to variable mass accretion rates.

The reason behind the variability of accretion is not understood.
\citet{turner10} proposed that changes in the disk magnetic flux coupled
with changes in the X-ray luminosity can lead to substantial changes in
the mass accretion rates of typical TTS disks.  However, this applies to 
accretion flows onto the star.  For the inner disk,  accretion variability
could be linked to the formation mechanism behind cavities in
disks, namely planets. One can speculate that changes in
mass accretion rate could be due to planetary companions which alter the
accretion flow in the inner disk regions, eventually getting onto the star. \citet{lubow06} and \citet{zhu10} demonstrate that planets will affect the mass accretion rate
into the inner disk.  
It is possible that this could occur on the timescales seen here given that the 1--3 yr variability observed in our sample corresponds to orbital timescales of 1--2 AU, plausible
radii for planets to be located.

\subsubsection{Disk Warps} \label{diskwarps}

The changes seen in the inner disk could be due to warps. To explain the
variability seen in the transitional disk of LRLL 13,
\citet{muzerolle09} proposed that the variability was due to dynamical
changes in the inner disk, particularly in the form of disk warps.
\citet{flaherty10} showed that the seesaw-like variability observed
could be explained by models of a disk with an inner warp which
leads the height of  the inner disk to change with time. Such warps
could be due to the presence of multiple planets in the disk. While a
disk would damp the eccentricity of a single planet, multiple planets
would be able to maintain eccentric orbits which would induce
modulations that would effect the inner disk edge \citep{dangelo06},
leading to the change in the height of the inner wall needed to
reproduce the observations of pre-transitional disks.

Warps caused by planets could account for the timescales of the flux
changes seen in our sample. Variability on timescales of 1--3 yr
corresponds to orbital timescales of 1--2 AU and 1 wk timescales
correspond to 0.07~AU. Radii of 1--2~AU are plausible locations for
planetary companions.  A radius of 0.07~AU is comparable to the dust
destruction radius.   Many ``hot Jupiters'' are known to exist at radii
$<$0.1~AU \citep{marcy05}, comparable to or within the magnetospheric
radius of their host stars and well within the dust sublimation radius,
most likely reaching their current positions via migration
\citep{rice08}.

\subsection{The Composition of Silicate Dust in the Disk}

The formation of crystalline silicates requires high temperatures
\citep{fabian01}. One proposed mechanism for the formation of
crystalline silicates is accretion heating in the innermost disk,
close to the dust sublimation radius \citep[e.g.][]{gail01}. However, in
our objects crystalline silicates can be seen in the outer wall, which
is located at radii much further than the dust sublimation radius. In
several other studies, it is also found that crystalline silicates are
in the outer regions of the disk \citep{bouwman08,watson09,sargent09,olofsson09,olofsson10}.
This suggests that large-scale radial mixing is necessary to transport
the crystalline silicates that form near the dust destruction radius out
to larger radii in the disk  \citep[e.g.][]{boss04, gail04, keller04, ciesla07},
but this must occur before the hole or gap forms.  Alternatively,
crystalline silicates can form in the inner disk due to heating from shocks
in the disk created by planets \citep{desch05, boss05, bouwman08}.
In this case, the crystalline silicates would still need to be transported to 
the outer disk.
Changes in the accretion rate or stellar luminosity could possibly lead to 
the formation of crystalline silicates in the surface layers of the disk as proposed in the
case of EX Lupi  \citep{abraham09}.
Or as noted
by \citet{espaillat07b}, local processing may be due to collisions of
larger bodies that produce small grains heated to temperatures high
enough to create crystals.

One interesting by-product of this study is the possibility to explore
how or if the silicate composition is linked to the variability seen in
the sample. Indeed, we find that the four disks with 1~wk
variability contain the highest amounts of
crystalline silicates in the sample.   In UX~Tau~A, ISO~52, T35, and
T56, the outer wall is composed of $\sim$19$\%$, 20$\%$, 60$\%$, and 18$\%$
crystalline silicates, respectively.  (The next highest fraction is
$\sim$13$\%$, found in SZ~Cha.)  The optically thin regions of ISO~52 and T56 are
$\sim$20$\%$ and $\sim$25$\%$
crystalline silicates, respectively, higher than any other disks in the sample by a factor of at least 2. 
In larger studies of samples
focusing on low-mass stars, there is very little correlation between the crystalline silicate mass fraction and any
stellar or disk property aside from the positive correlations with other crystalline silicate abundances
and the amount of dust settling in the disk \citep{sargent09,watson09}.  

The trend between crystalline mass fraction and short timescale variability
points to a link behind the underlying cause of the silicate composition
of the disk and the seesaw behavior observed. We propose that this link
is planets.  Planets can instigate warps, shocks, and collisions in the
disk which can lead to both changes in the height of the inner disk wall
and a higher abundance of crystalline silicates. 
Planets can also lead to recurring changes in the disk.
In T56 and T35, the variability observed at $\lambda >$ 20~{\micron} appears
to oscillate between a maximum and minimum flux.  For T56, the flux at these wavelengths
is at a minimum in the GTO epoch, it increases in the GO1 epoch, and in the GO2 epoch it
decreases back to the same flux observed in the GTO epoch.
In T35, the GTO and GO1 spectra are both at the maximum and the GO2 spectrum
is lower.
Such changes point to a periodic origin, such as a planetary orbit.
Since planets are
likely present in most of the disks in this sample, it would seem that
these four disks with high amounts of crystalline silicates either have more planets or that the presence of hot
Jupiters (see {\S}~\ref{diskwarps}) is significant.

\section{Summary \& Conclusions} \label{sed:sum}

In this work we see various types of variability on 3-4 year timescales
and in some cases we see variability on 1 week timescales. The dominant
type of variability observed can be classified as seesaw-like behavior,
whereby the emission at shorter wavelengths varies inversely with the
emission at longer wavelengths.

We attempted to understand the origin of the variability in pre-transitional
and transitional disks by modeling the overall
SEDs at different epochs.
For many of the pre-transitional disks
we find that the variability can be explained by changing the height of
the inner disk wall and hence the shadow on the outer disk.  Typically, the
height of the wall varies by $\sim$20$\%$.   We also
perform SED model fitting for the transitional disks GM~Aur and CS~Cha.
To fit the variability of GM~Aur, we vary the amount of
optically thin dust in the hole and the height of the
outer wall.  In CS~Cha, only
the 10~{\micron} silicate emission changes between epochs and so we only alter the amount of dust in the optically thin region. 
The transitional disks DM~Tau and T25 are the
only two disks in the sample which display no variability.  These disks' inner regions do not contain discernible amounts of dust. 

We propose that planets are responsible for the changes observed in
our sample.
Overall, it seems that most of the variability seen is due to material
in the inner disk casting a shadow on the outer disk.  The height of the
inner wall can vary due to disk warps caused by planets in the
disk. We can also link the silicate dust
compositions found in this work to the presence of planets.  
We find that crystalline silicates are
common in the outer disks of our objects, too far from the central star
to be explained by most crystallization mechanisms.  In addition, the four disks in 
our sample which have the highest crystalline silicate mass fractions
vary on 1~wk timescales.  In two of these four disks, we see periodic changes
in the infrared emission.  
Planets can cause shocks and collisions which can heat the dust to
high enough temperatures to crystallize the dust.  Planets
can also lead to short timescale and periodic variability. 
Follow-up variability studies conducted with the {\it James Webb Space Telescope} will 
give us the simultaneous, multi-wavelength data needed to 
test if the variability observed in our sample is periodic as well as the sensitivity to
significantly expand the sample size.

 \acknowledgments{
 We thank the referee for a constructive and thorough report.
 We thank Lee Hartmann for providing comments on the manuscript
and Steve Lubow for useful discussions.
C.~E.~was supported by the National Science Foundation under Award No. 0901947.  
 E.F. was supported by NASA through the Spitzer Space Telescope 
Fellowship Program, through a contract issued by JPL/Caltech 
under a contract with NASA.
P.~D.~acknowledges a grant from PAPIIT-DGAPA UNAM.
 E.~N. acknowledges a postdoctoral grant from CONACyT.
N.~C.~acknowledges support from NASA Origins Grant NNX08AH94G. 
}

\appendix

(Objects here are organized in the order in which they appear in Figures~10 through~17.)

\subsubsection{LkCa~15} \label{model:lkca15}

LkCa~15 is a pre-transitional disk in the Taurus cloud, previously
identified and modeled in detail by \citet{espaillat07b,espaillat10}.
The large cleared region in this disk has been confirmed with millimeter
imaging  \citep[Andrews et al, in prep;][]{pietu06}.  Here we take the model presented in
previous works and modify it to better fit the silicate dust features in
the SED and fit the observed variability. Figure~\ref{figlkca15} (top
left) shows the three {\it Spitzer} IRS spectra assembled for this
study.  The GO1 and GO2 spectra agree with one another.  The GTO
spectrum is higher than the GO spectra at $\lambda <$8~{\micron} and
lower at $\lambda >$15~{\micron}.  The 10~{\micron} silicate emission
remains the same in all three spectra.
Two additional spectra of LkCa~15 were taken two days apart in November 2008 by GO Program 
40338 (Nov 5; PI: Najita) using the SH module (10--19~{\micron}) and GTO Program 50498 (Nov 7; PI: Houck) using the LH module (19--37~{\micron}).
We find that the SH data agree with our GTO and GO data at ${\lambda}<$15~{\micron} and our
GO data at 15--19~{\micron}; the LH data agree with our GTO data.
This could suggest that 
the type of variability seen in LkCa~15 occurs on timescales as short as 2 days. However,
this is speculative and simultaneous data at shorter and longer wavelengths would be needed to test this.

Before trying to explain the variability, we first modeled the GTO
spectrum in detail in order to better fit the silicate dust features.
The composition of the silicate dust in the outer wall and disk that
gave the best fit to the IRS spectrum is listed in
Table~\ref{tab:silwall}.  The size of the gap is 39~AU. We note that to
better fit the upward slope of the IRS spectrum beyond
$\sim$30~{\micron} we used an ice abundance of 0.0012.  The abundance of
troilite and organics given in {\S}~\ref{sec:moddustopa} was not changed. 
As stated in {\S}~\ref{sec:moddustopa}, the inner wall is made up of olivine
silicates. The optically thin dust region spans 15~AU in the disk and
contains ISM-sized dust \citep[i.e. $a_{min}$=0.005~{\micron} and
$a_{max}$=0.25~{\micron};][]{mathis77}. Within 1~AU the dust is 32$\%$
organics,  13$\%$ troilite, and  55$\%$ silicates.  Between 1--15~AU the
dust is made up of 12$\%$ organics,  9$\%$ troilite, and  79$\%$
silicates. The compositions of the silicates in these regions are listed
in Table~\ref{tab:silthin}.  We found that having both olivines and
pyroxenes was necessary to fit the SED. The olivines in the inner 1~AU
contribute mostly to the 10~{\micron} silicate emission.  If olivines
were located further than 1~AU in the optically thin region, they would
produce too much 20~{\micron} silicate emission.  Therefore, we used
pyroxenes in the outer radii of the optically thin region to fit the
20~{\micron} silicate emission feature. The optically thin dust region
contains 2$\times$10$^{-9}$ M$_{\sun}$ of small dust.

We can fit the variability observed by changing the height of the inner
wall (Figure~\ref{figlkca15}, top left; GTO fit: solid line; GO1$/$GO2
fit: broken line).  In the GTO spectrum, the inner wall height is
0.009~AU and for the GO spectra it is 0.007~AU, corresponding to a 22$\%$ decrease in
the height of the wall.  The small change in the
inner wall is enough to cause the difference in flux observed at the
longer wavelengths. When the inner wall is higher (in
the GTO case), there is a larger shadow on the outer wall and hence the
emission seen from the outer wall is less and the IRS spectrum is lower
than the GO spectra.  In the GO case, the inner wall is lower and so the
shadow on the outer wall is smaller. Therefore, we see more emission and
the flux of the GO spectra longwards of 20~{\micron} is higher.  The
10~{\micron} silicate emission does not change. This emission is
dominated by the small dust in the optically thin region.

In the top right panel of Figure~\ref{figlkca15} we present the best fit
model to the GTO IRS spectrum and the broad-band SED. The disk
properties of LkCa~15 are listed in Table~\ref{tab:disk}. The inner wall
has a temperature of 1400~K and has a maximum grain size of
1~{\micron} following the best-fit found by fitting near-IR spectra from
1--5~{\micron} by \citet{espaillat10}.  The inner wall is located at a
radius of 0.15~AU. The height of the inner wall varies as outlined
above.  The outer wall has a temperature of 110~K, corresponding to a
radius of 39~AU.  The height of the outer wall is about 5~AU. The outer
disk has $\alpha$=0.0005, $\epsilon$=0.001, and a mass of
0.1~M$_{\sun}$. The differences between the parameters presented here
and those of \citet{espaillat10}, especially the $\sim$20~AU difference in radius, are due to the different dust
composition adopted.  See the discussion in that work for details on how
different dust compositions lead to differing fits and hence derived
disk properties.

We note that a pre-transitional disk structure for LkCa 15 has also been
independently found through SED modeling by \citet{mulders10}. However,
those authors assume the star is a point source and so the shadowing of
the outer disk is not properly calculated.  \citet{espaillat10} and this
work take into account that the star is a finite source and the ensuing
effect on the heating of the outer disk.

\subsubsection{SZ~Cha} \label{model:szcha}

SZ~Cha was first identified as a pre-transitional disk candidate by
\citet{kim09}. Here we present the first detailed model of this object. 
The GO1 and GO2 IRS spectra of SZ~Cha agree with one another
(Figure~\ref{figszcha}, middle left). However, the GO spectra are higher
than the GTO spectra at $\lambda <$8~{\micron} and lower at $\lambda
>$15~{\micron}.  The 10~{\micron} silicate emission is similar in all
three spectra.

One additional SH spectra of SZ~Cha was taken in GO Program 
40247 (PI: Calvet) on August 17, 2008.  We find that this spectrum agrees with
our GTO and GO spectra between 10--13~{\micron}.  However,
between 13--19~{\micron} the flux in the SH spectrum is lower than both
the GTO and GO observations.  In particular, the flux is about 15$\%$ lower
in the SH observations than our GO observations.  We do not have simultaneous
data at shorter wavelengths and so we do not model the SH data here.  However,
these SH observations support that the 10~{\micron} silicate emission feature does not vary while
the emission at longer wavelengths does.  Furthermore, the SH data were taken two months after our GO observations, showing that variability in SZ~Cha occurs on shorter timescales than the 3 year timescale
probed by our GTO and GO observations.

We fit the IRS spectrum with a 18~AU gap and the silicate dust mixture
listed in Table~\ref{tab:silwall}.  The optically thin dust extends up
to 1~AU in the disk and contains 0.005--3~{\micron} dust composed of 12$\%$
organics, 9$\%$ troilite, and 79$\%$ silicates.  The silicate
composition can be found in Table~\ref{tab:silthin}. The gap contains
2$\times$10$^{-12}$ M$_{\sun}$ of small optically thin dust.

We fit the observed variability by adjusting the height of the inner
optically thick disk's
wall (Figure~\ref{figszcha}, middle left; GTO fit: solid line; GO1$/$GO2
fit: broken line).  In the GTO spectrum, the inner wall height is
0.006~AU and for the GO spectra it increases by 33$\%$ to 0.008~AU.  This change in
the inner wall is enough to cause the difference of flux observed at the
longer wavelengths.  The 10~{\micron} silicate emission, which is
dominated by the small dust in the optically thin region, does not
change significantly. In the middle right panel of Figure~\ref{figszcha}
we present the best fit model to the GTO IRS spectrum and the broad-band
SED. Model parameters are listed in Table~\ref{tab:disk}.

\subsubsection{UX~Tau~A} \label{sec:uxtaua}

UX~Tau~A was previously identified as a pre-transitional disk and
modeled by \citet{espaillat07b,espaillat10}. Millimeter imaging has
confirmed that there is a large cavity in this object (Andrews et al, in
prep). As in the case of LkCa~15, we re-model UX~Tau~A, paying
particular attention to fitting its silicate emission features in order
to better reproduce the variability.  

\begin{figure}
\epsscale{1}
\plotone{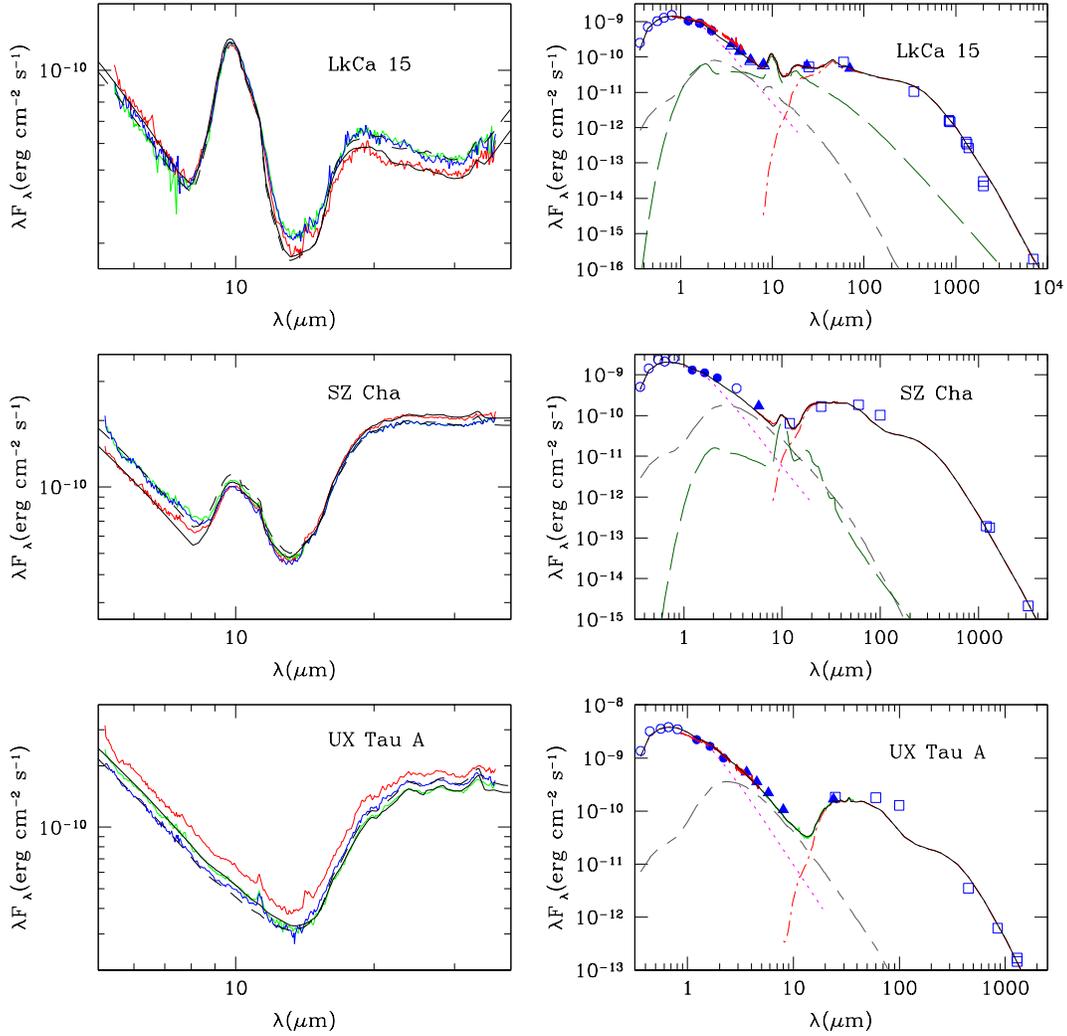}
\caption[]{SEDs and disk models of LkCa~15, SZ~Cha, and UX~Tau~A. 
The left panels show the IRS GTO (red), GO1 (green), and GO2 (blue)
spectra for the target plotted with the best fit models (solid and
broken black lines; refer to the Appendix for more details).
The right panels show the best fit model to the GTO IRS spectrum along
with the rest of the broad band SED. In the case of UX~Tau~A, the best
fit to the GO1 spectrum is used. SpeX spectra first presented in \citet{espaillat10} are also plotted
in the cases of LkCa 15 and UX Tau A.  
Open circles correspond to ground-based (UBVRIL) photometry; for Taurus these data are
taken from \citet{kh95} and for Chamaeleon we use photometry from \citet{gauvin92}.
Closed circles are 2MASS photometry from \citet{skrutskie06}.
Triangles are {\it Spitzer} photometry (see Table~\ref{tab:spitzer} for more details).
Squares correspond to IRAS and millimeter fluxes.
IRAS data are take from \citet{weaver92}.
Millimeter data for LkCa 15 are from \citet{andrews05, espaillat07b, rodmann06, kitamura02}. 
Millimeter fluxes for SZ Cha were measured by \citet{henning93} and
\citet{lommen09b}. 
UX Tau A's millimeter data are from
\citet{andrews05} and \citet{beckwith90}.
The observations have be dereddened using the extinctions listed
in Table~\ref{tab:disk}. Separate model components are the stellar photosphere
\citep[magenta dotted line;][]{kh95}, the inner disk (gray
short-long-dash), the outer disk (red dot-short-dash), and the optically
thin small dust located within the inner disk (green long-dash). 
(A color version of this figure is available in the online journal.)
}
\label{figlkca15}
\label{figuxtaua}
\label{figszcha}
\end{figure}

The variability seen in UX~Tau~A is similar in some respects to that seen 
in LkCa~15, but very different in others.  
The GO spectra display variability like that of LkCa~15, but without
the presence of a discernible 10~{\micron} silicate emission feature (Figure~\ref{figuxtaua}, bottom left).
Shortwards of $\sim$18~{\micron}, the GO1 spectrum (green) is higher
than the GO2 spectrum (blue).  At longer wavelengths the behavior
switches, with the GO2 spectrum now higher than the GO1 spectrum.

The GTO spectrum of UX~Tau~A is higher than both of
the GO spectra at all wavelengths (Figure~\ref{figuxtaua}, bottom left).
The overall decrease in the emission between our GTO and GO spectra
suggests that the luminosity at the bands at which the disk absorbs stellar radiation has decreased between the two
epochs.  Without simultaneous optical data, we cannot test this.
Alternatively, the mass accretion rate of the disk could have decreased over time.
Two LH spectra were taken by GO Program 
2300 (PI: Carr) on March 14, 2005 and GTO Program 50498 (PI: Houck) on November 6, 2008.
Both are consistent within the observational uncertainties with our GTO spectrum which was
taken in February 2004.  
However, we do not have data at $<$19~{\microns} and cannot discern how$/$if the SED changes at
shorter wavelengths.

One SH spectrum and one LH spectrum was taken in GO Program 30300 (PI: Najita)
on March 22, 2007, between the times at which our GTO observations and GO observations were taken.  The SH spectrum agrees with our GO2 spectrum between 10 and 13~{\microns}, but is substantially lower at longer wavelengths (Figure~\ref{uxshlh}).  The LH spectrum is also lower than our GTO
and GO data.  This suggests that at ${\lambda}>$13~{\microns}
UX Tau A returns to the higher fluxes seen in our GTO and GO data periodically, possibly within about two years.  Again, we would need simultaneous data at ${\lambda}<$10~{\micron} to come to a firmer
conclusion.

We note that UX Tau A has one companion at 2.5$^\prime$$^\prime$ (UX Tau C) and one at 5.6$^\prime$$^\prime$ (UX Tau B). In November
2002, UX Tau B was fainter than A by a factor of 12 at
11.8~{\micron}, while UX Tau C was fainter than A by a factor
of 60 \citep{mccabe06}.  
UX Tau B and C entered all the IRS slits except in the case of the SL slit where UX Tau B did not enter the slit.
UX Tau B and C do not have
a near-infrared excess and are not accreting \citep{white01} and so we can conclude that they
most likely do not have disks.  Therefore, they would not contribute at wavelengths $>$13{\micron} where we see a large change in flux in Figure~\ref{uxshlh}.  Therefore, this variability is probably not
due to large changes in the fluxes of these companions.

In Table~\ref{tab:silwall} we list the silicate dust composition that
best explains the crystalline silicate features seen in the IRS spectrum
beyond 20~{\micron}. To better reproduce the spectrum at longer
wavelengths we removed ice from the outer wall and disk.  Our model of
UX~Tau~A includes no optically thin dust within the gap, which is 30~AU
in radius.

As in the case of LkCa~15, we can fit the variability observed
between the GO spectra by varying the height of the inner wall
(Figure~\ref{figuxtaua}, bottom left; GO1 fit: solid line; GO2 fit:
broken line). In the GO1 spectrum the height of the wall is 0.006~AU and it
decreases by 17$\%$ to 0.005~AU in the GO2 spectrum. This difference in the inner
wall height (and hence the amount of shadowing on the outer disk wall)
is enough to cause the observed flux change at the longer wavelengths. 
Here we adopt the temperature for the inner wall found by
\citet[][1550~K]{espaillat10} based on fitting near-IR spectra.
However, we find that we can fit the SHLH spectrum (Figure~\ref{uxshlh}) with an inner wall
temperature of 1800~K (corresponding to a radius of 0.1~AU) and a height of 0.0053~AU. 
Moving the wall closer to the star produces a larger shadow on the outer wall, and reproduces the
SHLH spectrum well.  However, we do not have simultaneous data at shorter wavelengths and
hence cannot test if this wall fits the SED at ${\lambda}<$10~{\micron} at the time the SHLH spectrum was taken.  Further observations
are needed to test if the temperature of the inner wall of UX~Tau~A varies with time.

The bottom right panel of Figure~\ref{figuxtaua} has the best fit model
to the GO1 IRS spectrum and the broad-band SED.  Relevant disk
properties are listed in Table~\ref{tab:disk}. The differences between
the parameters presented here and those of \citet{espaillat10} are
substantial, particularly the 40~AU difference in the radius of the gap.
 Again, this is due to the different dust composition adopted. 
Including crystalline silicates (forsterite, enstatite, silica) has a
substantial effect on the opacities and hence disk properties.

\subsubsection{RY~Tau} \label{model:rytau}

Analysis by \citet{furlan09} identified RY~Tau as an outlier with
respect to its EW(10~{\micron}).  RY~Tau had much more silicate emission than could be explained
by the full disk model grid used in that study.  As a part of this
variability study, we attempted to fit the disk of RY~Tau with a full
disk model and were unsuccessful. It is not possible to simultaneously
fit the existing millimeter and IRS data, even within the uncertainties
of the observations.  The strong 10~{\micron} silicate emission cannot
be reproduced by a typical full disk model and could be a sign that
RY~Tau is a pre-transitional disk with a small gap that contains some
small optically thin dust, as suggested by \citet{espaillat09} and
\citet{furlan09}.    Submillimeter imaging by \citet{isella10} supports
that there is an inner cavity in this disk.

\begin{figure}
\epsscale{1}
\plotone{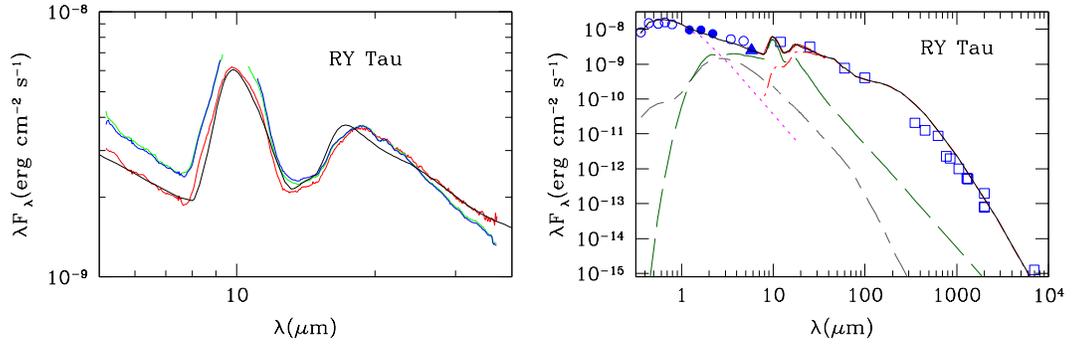}
\vskip -4in
\caption[]{SED and disk model of RY~Tau.  
The left panels show the IRS GTO (red), GO1 (green), and GO2 (blue)
spectra for the target plotted with the best fit models (solid and
broken black lines; refer to the Appendix for more details).
Plotted data are the 
same as described in Figure~\ref{figlkca15}. Note that for RY~Tau the region 
around 10~{\micron} in the GO spectra has been removed due to saturation.
Millimeter data for RY Tau are from 
\citet{andrews05} and
\citet{andrews07}. 
(A color version of this figure is available in the online journal.)
}
\label{figrytau}
\end{figure}

\begin{figure}
\epsscale{1}
\plotone{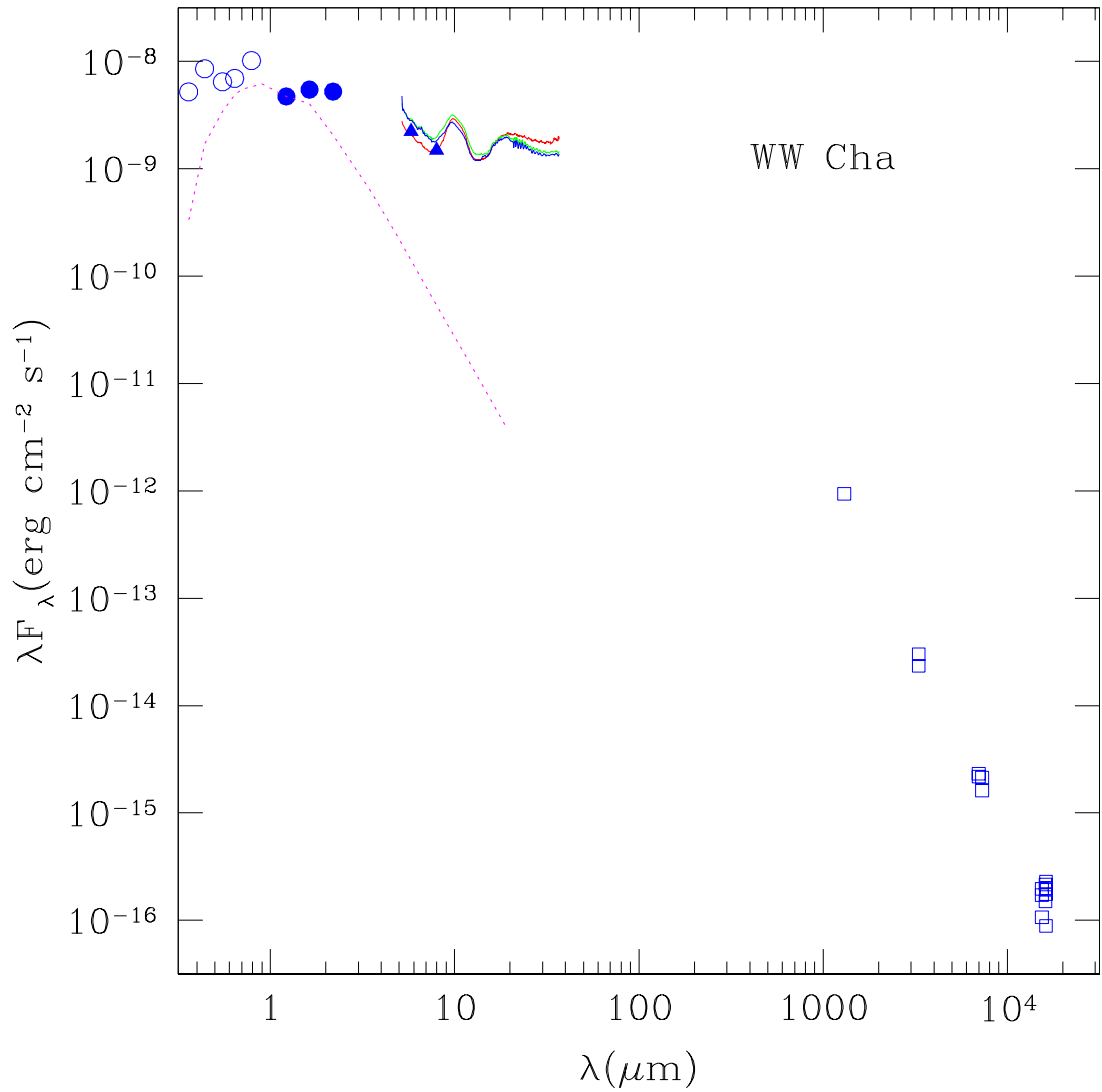}
\caption[]{Broad-band SED of WW~Cha including
the IRS GTO (red), GO1 (green), and GO2 (blue)
spectra.  See the Appendix for further discussion. 
Refer to the caption of Figure~\ref{figlkca15}
for more details on the plotted data.
Millimeter fluxes are taken from the literature \citep{henning93, lommen07, lommen09}.
(A color version of this figure is available in the online journal.)
}
\label{figwwcha}
\end{figure}

Figure~\ref{figrytau} (left) shows the three {\it Spitzer} IRS spectra
assembled for this study.  The GO spectra agree with one another.  The
GTO observation was mispointed, and so we cannot confidently constrain
the amount of variability between the GTO and GO spectra nor the pivot
wavelength.  In addition, the 10~{\micron} region in the GO spectra were
saturated and we cannot reliably fit the emission from the inner disk of
RY~Tau. 
Two additional high-resolution spectra of RY~Tau were taken in GO Program
40113 (SH; PI: Lahuis) and Program 50641 (LH; PI: Carr), both on October 2, 2008.  We find that the SH spectrum agrees with
our GO spectra shortwards of 15~{\micron}.  However,
at longer wavelengths the flux is consistent with both
the GTO and GO observations within the observational uncertainties.   The LH spectrum is also consistent with
both the GTO and GO observations.

We found that we could fit the GTO IRS spectrum with an
18~AU gap (Figure~\ref{figrytau}, both panels) and a silicate dust
mixture in the outer wall and disk consisting of predominantly olivine
silicates (Table~\ref{tab:silwall}).  The optically thin dust in the gap
extends up to 1~AU in the disk and contains 3$\times$10$^{-10}$
M$_{\sun}$ of 0.005--3~{\micron} dust composed of 19$\%$ organics, 15$\%$
troilite, and 66$\%$ silicates.  The silicates in the optically thin dust region have the same composition
as that in the wall and outer disk (Table~\ref{tab:silthin}). Disk
properties are listed in Table~\ref{tab:disk}.  Here we adopt the same
mass accretion rate as \citet[][Table~\ref{tab:stellar}]{calvet04}.

\subsubsection{WW~Cha} \label{wwcha}

WW~Cha displays the same type of variability as seen in LkCa~15, SZ Cha,
and UX~Tau~A (Figure~\ref{figwwcha}).  The GO1 and GO2 IRS spectra of
WW~Cha agree with one another and the GO spectra are higher than the GTO
spectra at $\lambda <$8~{\micron} and lower at $\lambda >$15~{\micron}.
However, because the GO observations were mispointed, we cannot
confidently constrain the amount of variability nor the pivot wavelength.
 Clearly there is true variability as we would have expected the spectra
to have lost flux overall due to the mispointing, but at the shorter
wavelengths the GO spectra are higher than the GTO spectra.
We note that there are no additional {\it Spitzer} IRS spectra of WW~Cha
available in the archive.

We do not model this object because it was mispointed in the GO observations and we do not have a reliable mass
accretion rate estimate.  Based on the SED it appears that the optical
photometry is unreliable.  On the other hand, it could be that the
accretion rate of WW~Cha is very high, which would lead to a substantial
excess above the stellar photosphere $<$1~{\micron}.  If the photometry
is accurate, we derive a mass accretion rate of 8$\times$10$^{-7}$
{\msun} yr$^{-1}$, making WW~Cha a high accretor like DR Tau.  D'Alessio
et al. (in prep) discuss the considerations that must be taken into
account when modeling high accretors.  Because of these reasons
and the fact that the GO spectra were mispointed, we do not
attempt to model WW~Cha in this work.   However, the GTO IRS spectrum of
WW~Cha is similar in morphology to that of LkCa~15, suggesting that it
is a pre-transitional disk as well.

\subsubsection{ISO~52}  \label{iso52}

\begin{figure}
\epsscale{1}
\plotone{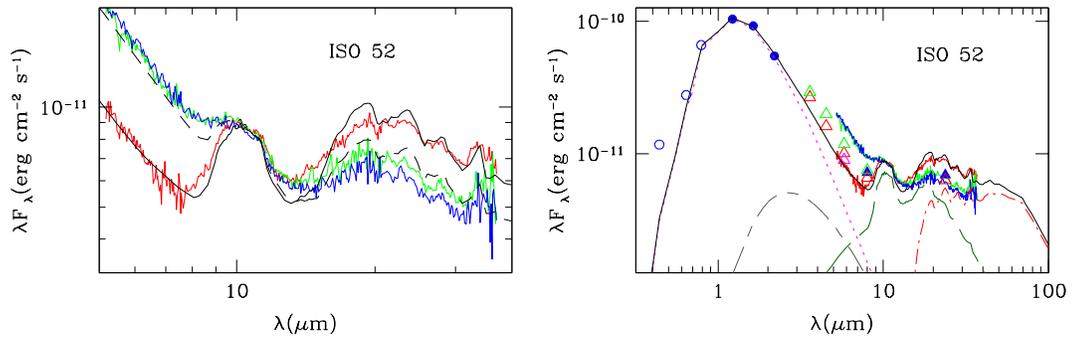}
\vskip -4in
\caption[]{Broad-band SED of ISO~52.  
The left panels show the IRS GTO (red), GO1 (green), and GO2 (blue)
spectra for the target plotted with the best fit models (solid and
broken black lines; refer to the Appendix for more details).
Refer to the caption of Figure~\ref{figlkca15}
for more details on the plotted data.   B and R
photometry (open circles) are from \citet{monet98}. I-band photometry
(open circle) is from DENIS.
(A color version of this figure is available in the online journal.)
}
\label{figiso52}
\end{figure}

ISO~52 is located in Chamaeleon and its SED is presented in
Figure~\ref{figiso52}. ISO~52 displays seesaw-like behavior; however, it
is very different from any of the other disks in the sample.  The GO
spectra (which agree with one another) are substantially higher than the GTO spectra
at $\lambda <$9~{\micron} and lower at $\lambda >$12~{\micron}.  The
10~{\micron} silicate emission feature almost disappears in the GO
spectra due to the increasing continuum in the 5--8~{\micron} region. 
There are no additional {\it Spitzer} IRS spectra of ISO~52
available in the archive.

As in the case of WW~Cha, we do not
have a mass accretion rate estimate for ISO~52;  there is no U-band photometry for
this object.  However, for the purposes of reproducing the observed variability
we modeled the SED assuming a mass accretion rate of 1$\times$10$^{-8}$ {\msun} yr$^{-1}$.
We fit the IRS spectrum with a 10~AU gap and the silicate dust mixture
listed in Table~\ref{tab:silwall}.  The optically thin dust extends up
to 1~AU in the disk and contains 2$\times$10$^{-12}$ M$_{\sun}$ of 0.005--5~{\micron} dust composed of 42$\%$
organics, 11$\%$ troilite, and 47$\%$ silicates (Table~\ref{tab:silthin}). 

We fit the observed variability by adjusting the height of the inner
optically thick disk's
wall (Figure~\ref{figiso52}, left; GTO fit: solid line; GO1$/$GO2
fit: broken line).  In the GTO spectrum, the inner wall height is
0.0006~AU and for the GO spectra it is 0.003~AU.  This corresponds to
an increase of 400$\%$ in the wall height, the largest change by far in any of the objects
in the sample.  To fit the GO observations, the optically thin dust region has to be extended to
2~AU and contains three times more dust to account for the emission at 20~{\micron} since the contribution from the outer wall at these
wavelengths is significantly less relative to its contribution in the
GTO epoch. 

In the right panel of Figure~\ref{figiso52}
we present the best fit model to the GTO IRS spectrum and the broad-band
SED. We stress that the mass accretion rate of ISO~52 is not known.
As such we do not list its disk model parameters in Table~\ref{tab:disk}
along with disks where the mass accretion rate is known and hence the
disk parameters are better constrained.  Here we briefly summarize the disk
parameters obtained by adopting a mass accretion rate of 1$\times$10$^{-8}$ {\msun} yr$^{-1}$.
The inner wall is located at a temperature of 1400~K with a radius of 0.05~AU.  It has a maximum grain size of
10~{\micron} and its height varies as outlined above.  The outer wall has a temperature of 130~K, corresponding to a
radius of 10~AU.  The height of the outer wall is about 1~AU. There is no millimeter information for this
object and so we do not model the outer disk.

\subsubsection{CR~Cha} \label{chx3}

\begin{figure}
\epsscale{1}
\plotone{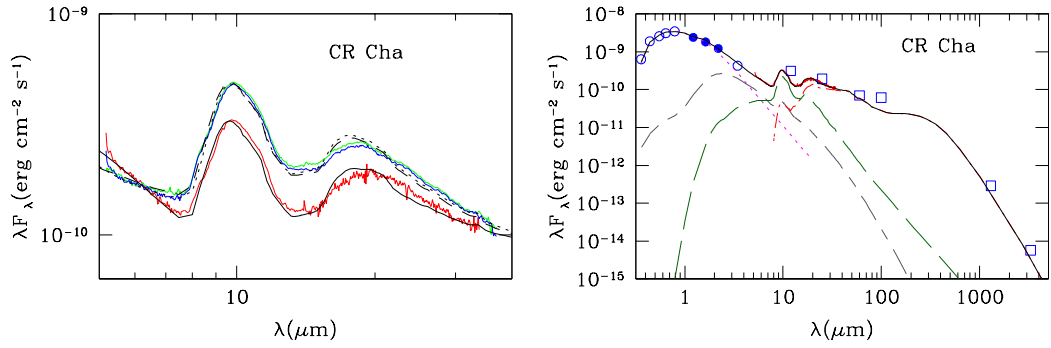}
\vskip -4in
\caption[]{SED and disk model of of CR~Cha.  
The left panels show the IRS GTO (red), GO1 (green), and GO2 (blue)
spectra for the target plotted with the best fit models (solid and
broken black lines; refer to the Appendix for more details).
Refer to the caption of Figure~\ref{figlkca15}
for more details on the plotted data.
Millimeter fluxes are from \citet{henning93} and
\citet{lommen07}.
(A color version of this figure is available in the online journal.)
}
\label{figchx3}
\end{figure}

Analysis by \citet{furlan09} identified CR~Cha as an outlier with
respect to its EW(10~{\micron}).  Here we find that it is not possible
to simultaneously fit the existing data with a full disk model, as in
the case of RY~Tau. Figure~\ref{figchx3} (left) shows the three {\it
Spitzer} IRS spectra assembled for this study.  The GO1 and GO2 spectra
agree with one another and are both higher than the GTO spectra at
$\lambda >$6~{\micron} and lower at $\lambda <$6~{\micron}.
We note that there are no additional {\it Spitzer} IRS spectra of CR~Cha
available in the archive.

From the substantial change in slope at $\sim$6~{\micron}, one could
infer that there is either a change in the nature of the emission from
optically thick to optically thin or a substantial change in the temperature of
the inner wall.  We found that we could fit the GTO
data with a pre-transitional disk model and the GO data with both a
transitional and pre-transitional disk model.  In all models the outer wall is located at
10~AU and is composed of the silicate mixture listed in
Table~\ref{tab:silwall}. 
We note that attempts to fit the variability of the SED with changes in the mass accretion rate
were unsuccessful.   High-resolution near-interferometric data of this object is necessary
to constrain how or if the inner disk structure responsible for the near-IR variability changes with time.

In the GTO fit (Figure~\ref{figchx3}, left,
solid line), there is an inner wall with a temperature of 1400~K, a radius of 0.26~AU, and a height
of 0.008~AU.  The outer wall has a height of 0.8~AU.
The optically thin dust in the gap
extends up to 1.5~AU and contains 1.8$\times$10$^{-11}$ M$_{\sun}$ of
0.005--10~{\micron} dust composed of 12$\%$ organics, 10$\%$ troilite, and 78$\%$
silicates.  The silicates in this region have the same composition as
that in the outer wall and disk (Table~\ref{tab:silthin}). 

In the GO pre-transitional disk model fit
(Figure~\ref{figchx3}, left, dotted line), the inner
wall has a temperature of 800~K, corresponding to a radius of 1~AU, and a height of 0.03~AU.   The outer wall has a height of 0.8~AU.  The optically thin region spans 1--1.5~AU, and contains 1.3$\times$10$^{-11}$ M$_{\sun}$ of
0.005--10~{\micron} dust  composed of
8$\%$ organics, 6$\%$ troilite, and 86$\%$ silicates.

In the GO transitional disk
model fit (Figure~\ref{figchx3}, left, dashed line), there is no inner
wall.  The optically thin region spans the same radii as in the GTO fit, but the
amount of dust has increased by a factor of two and is now composed of
17$\%$ organics, 13$\%$ troilite, and 70$\%$ silicates. The outer wall is
also the same as that in the GTO fit, except that the height of the outer
wall differs. In the GTO fit, which takes into account shadowing by the
inner wall, the height of the outer wall is 0.8~AU and in the GO fit,
where there is no shadowing it is 0.5~AU high.  This could imply that some
portion of the outer wall is shadowed in the GO fit, where there is only
optically thin dust in the inner disk. 

The outer disk
does not change between the epochs observed (Figure~\ref{figchx3},
right); properties are listed in Table~\ref{tab:disk}.

\subsubsection{IP~Tau} \label{model:iptau}

\begin{figure}
\epsscale{1}
\plotone{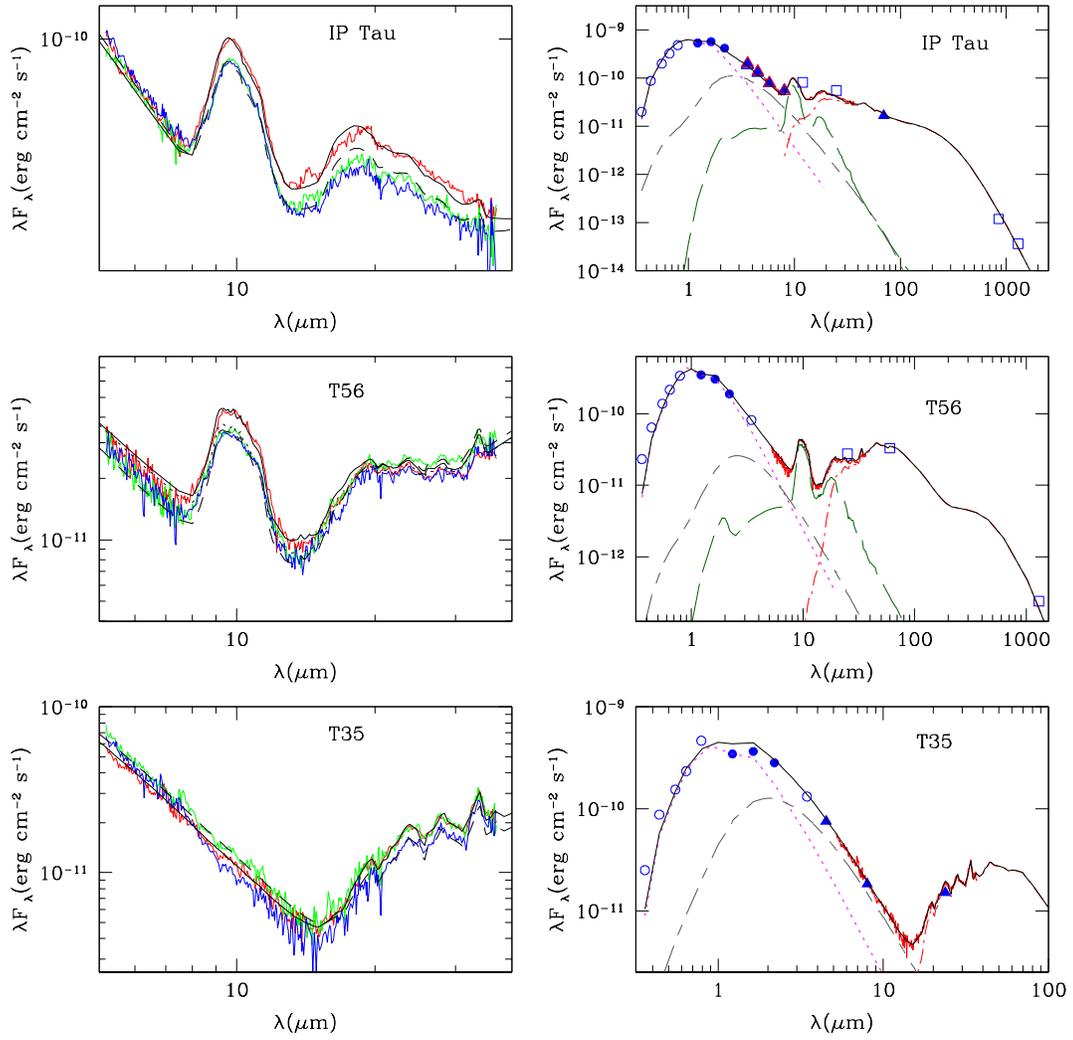}
\caption[]{SEDs and disk models of IP~Tau, T56, and T35.  
The left panels show the IRS GTO (red), GO1 (green), and GO2 (blue)
spectra for the target plotted with the best fit models (solid and
broken black lines; refer to the Appendix for more details).
Symbols, data references, and model lines are 
the same as noted in the caption of Figure~\ref{figlkca15}.
Millimeter data for IP Tau and T56 are from \citet{andrews05} and \citet{henning93}, respectively.
(A color version of this figure is available in the online journal.)
}
\label{figiptau}
\label{figsz45}
\label{figsz27}
\end{figure}

Analysis by \citet{furlan09} identified IP~Tau as an outlier with
respect to EW(10~{\micron}), as in the case of RY~Tau.  
 As a part of this work, we attempted to fit the disk of IP~Tau with a
full disk model and were unsuccessful. It is not possible to
simultaneously fit the existing millimeter and IRS data, even within the
uncertainties of the observations. Here we present the first detailed
model of this object, assuming it is a pre-transitional disk. The GO1
and GO2 IRS spectra of IP~Tau agree with one another within the
uncertainties, but variability occurs between the GTO and GO spectra
(Figure~\ref{figiptau}, top left). The variability seen in IP~Tau
differs from the variability discussed up to this point. The GTO spectra
are higher than the GO spectra at $\lambda >$8~{\micron}.  The
seesaw-like behavior seen in LkCa~15, UX~Tau~A, and SZ~Cha is not
conspicuous here. However, we attempt to reproduce the variability
assuming that the inner wall varies within the observational
uncertainties and are successful.
We note that two additional IRS spectra of IP~Tau were taken in GO Program
179 (PI: Evans) on February 8, 2005.  
We find that these SH and LH spectra are consistent with
the GTO data in this paper (taken in February 2004) within the observational uncertainties. 

The size of the gap in IP~Tau is very small, about 2~AU wide.  Details
of the silicate dust mixture used in the outer wall and disk can be
found in Table~\ref{tab:silwall}. The optically thin dust in the gap
extends up to less than 0.5~AU and contains 0.005--4~{\micron} dust composed of
17$\%$ organics, 13$\%$ troilite, and 70$\%$ silicates.  The silicates
have the same composition as that in the wall and outer disk.  The gap
contains 9$\times$10$^{-13}$ M$_{\sun}$ of small optically thin dust.

In Figure~\ref{figptd2err} there is no variability in IP Tau at the
short wavelengths, at least not outside of the uncertainties of the
observations.  However, IP Tau displays variability at longer
wavelengths similar to behavior observed in other objects in the sample
that could be explained by varying the height of the inner wall.
Therefore, we vary the height of the inner wall of IP Tau to fit the data within the
uncertainties of the observations.  For the GTO spectrum, the inner wall
height is 0.006~AU and for the GO spectra it is 0.007~AU i.e. the height of the
wall increases by 17$\%$
(Figure~\ref{figiptau}, top left; GTO fit: solid line; GO1$/$GO2 fit:
broken line).  This change in the inner wall reasonably reproduces the
difference of flux at the longer wavelengths. The 10~{\micron} silicate
emission also decreases between the GTO and GO observations.  This
emission is dominated by the small dust in the optically thin region
which may indicate that the amount of dust in this region is changing. 
We can reproduce the 10~{\micron} emission in the GO spectra by reducing
the amount of dust in the optically thin region by 20$\%$. The
contribution to the SED from the outer disk is shown in the top right
panel of Figure~\ref{figiptau}. We list the disk's properties in
Table~\ref{tab:disk}.

\subsubsection{T56}  \label{model:t56}

T56 is located in Chamaeleon and was first identified as a
pre-transitional disk candidate by \citet{kim09}. Here we present the
first detailed model of this object. The spectrum of T56 is noisy at
$\lambda <$8~{\micron}, but it seems to display the same type of pivot
variability as seen previously between the GTO and GO1 spectra
(Figure~\ref{figptd2err}). The GO1 and GO2 spectra agree in their
10~{\micron} silicate emission (Figure~\ref{fig1wkerr}), which is lower
than seen in the GTO spectrum. The GO2 and GTO IRS spectra of T56 agree
with one another beyond 20~{\micron} while the GO1 spectrum is higher at
these wavelengths.
There are no additional {\it Spitzer} IRS spectra of T56
available in the archive.

We fit the IRS spectrum with a 20~AU gap.  The silicate dust mixture in
the outer wall and disk is listed in Table~\ref{tab:silwall}.  The
optically thin dust region extends up to 1~AU in the gap and contains
1$\times$10$^{-11}$ M$_{\sun}$ of ISM-sized dust (in the GTO fit) composed of 23$\%$ organics, 14$\%$ troilite, and 63$\%$
silicates (see Table~\ref{tab:silthin} for silicate composition).

As noted above, at $\lambda <$8~{\micron} the SNR is poor and we cannot
confidently tell if there is variability at these short wavelengths.
However, we fit the observed variability by varying the height of the
inner wall within the uncertainties of the observations, as we did in
the case of IP~Tau.  In the GTO and GO2 spectra (Figure~\ref{figsz45},
middle left; GTO fit: solid line; GO2 fit: dotted line), the inner wall
height is 0.002~AU and for the GO1 spectra (Figure~\ref{figsz45}, middle
left; GO1 fit: dashed line) it is 0.001~AU, a 50$\%$ decrease.  The 10~{\micron} silicate
emission, which is dominated by the small dust in the optically thin
region, changes significantly and we reproduce this by decreasing the
amount of optically thin dust by 19$\%$ between the GTO and GO
observations.

In the middle right panel of Figure~\ref{figsz45} we present the best
fit model to the broad-band SED. Relevant disk properties are listed in
Table~\ref{tab:disk}. We note that the disk mass derived here is Toomre
unstable, i.e. Q$<$1 at about 100 AU. Given that we only have a single
millimeter data point to constrain the outer disk, this mass is poorly
constrained and more millimeter data is needed.

\subsubsection{T35} \label{sec:t35}

T35 is located in Chamaeleon and was first identified as a
pre-transitional disk candidate by \citet{kim09} and here we present the
first detailed model of this object. T35 displays the same type of
seesaw variability as seen in the previous objects.  However, it more
closely resembles UX~Tau~A, both in that it has no apparent 10~{\micron}
silicate emission and that it has strong crystalline silicate features. 
  As in the case of T56, discerning the variability is complicated at
the shorter wavelengths by poor SNR. At $\lambda <$10~{\micron}, we
cannot confidently tell if there is variability at these short
wavelengths. Interestingly, in the GO2 spectrum it seems that the slope
changes at 7~{\micron}. The GO1 and GTO IRS spectra of T35 agree with
one another beyond 20~{\micron} (Figure~\ref{figsz27}, bottom left).  
The GO2 spectrum is lower at these wavelengths.
One additional SH IRS spectrum of T35 was taken in GO Program
40247 (PI: Calvet) on October 12, 2008.  This SH spectrum is very noisy and consistent with
all three spectra presented in this paper within the observational uncertainties.  

We found that we could fit the IRS spectrum well with a gap of 15~AU and
the silicate dust mixture listed in Table~\ref{tab:silwall}
(Figure~\ref{figsz27}, bottom left; GTO$/$GO1 fit: solid line; GO2 fit:
broken line).  As in the case of UX~Tau~A, there is no optically thin
dust within the gap. At $\lambda <$10~{\micron}, the SNR is poor and we
cannot confidently tell if there is variability at these short
wavelengths. However, we fit the observed variability by varying the
height of the inner wall within the uncertainties of the observations. 
In the GTO and GO1 spectra, the inner wall height is 0.005~AU and for
the GO2 spectra it is 0.006~AU, a 20$\%$ increase in height, which leads to
the difference of flux at the longer wavelengths.  We
cannot reproduce the change in slope seen in the GO2 spectrum at
7~{\micron}.  One can speculate that this is somehow related to the fact
that in order to fit the SED we need a higher inner wall temperature
(1800~K) than used in the other disks in this sample.

In the bottom right panel of Figure~\ref{figsz27} we present the best
fit model to the broad-band SED.  Disk properties can be found in
Table~\ref{tab:disk}. We could only find a millimeter upper limit from
\citet{henning93} which is not useful in constraining the outer disk and
so we do not include modeling of the outer disk. However, we can still
reproduce the IRS emission beyond 20~{\micron} using only an outer wall.
 This reflects the point made in {\S}~\ref{sec:mod} that the outer
wall dominates the emission in this region of the IRS spectrum.

\subsubsection{GM~Aur}  \label{gmaur}

\begin{figure}
\epsscale{1}
\plotone{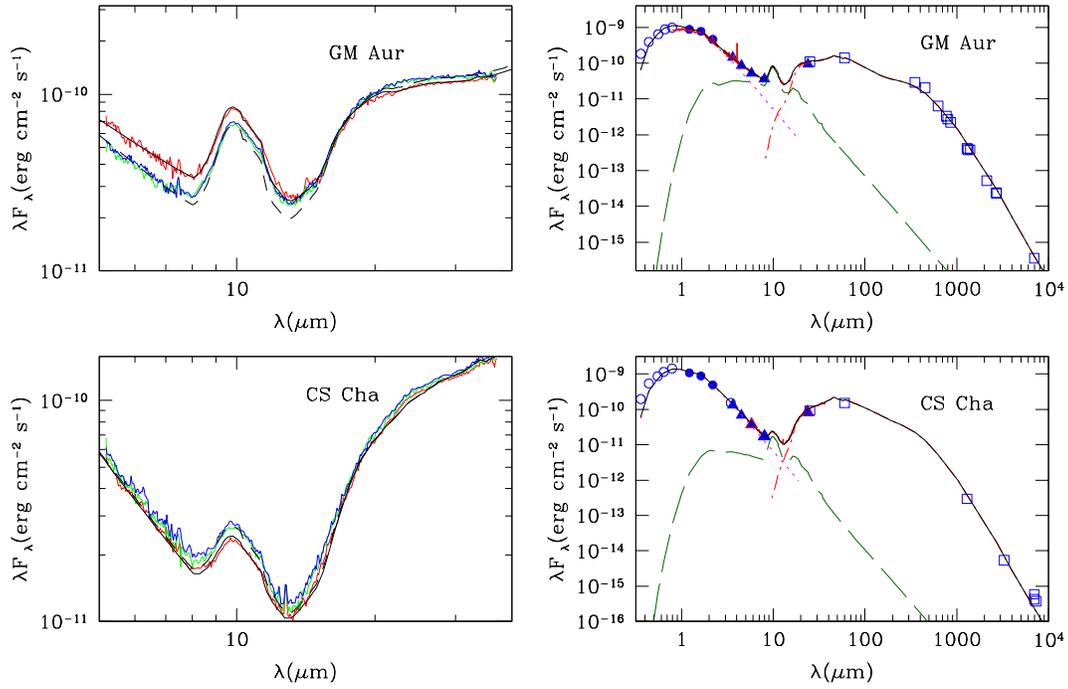}
\vskip -2in
\caption[]{SEDs and disk models of GM~Aur and CS~Cha.  
The left panels show the IRS GTO (red), GO1 (green), and GO2 (blue)
spectra for the target plotted with the best fit models (solid and
broken black lines; refer to the Appendix for more details).
Data are the 
same as listed in Figure~\ref{figlkca15}. SpeX spectra first presented in \citet{espaillat10} 
are also plotted in the case of GM Aur. Millimeter data for GM Aur were obtained from \citet{andrews05,weintraub89,beckwith91,dutrey98,koerner93,looney00,rodmann06,
hughes09}. 
Millimeter fluxes for CS Cha are from \citet{weaver92,
henning93,lommen07,lommen09}. 
(A color version of this figure is available in the online journal.)
}
\label{figgmaur}
\label{figcscha}
\end{figure}

GM~Aur was previously modeled in detail by \citet{calvet05}.  They found
it had an inner disk hole of $\sim$20~AU that contained some small
optically thin dust grains.  Veiling analysis of near-infrared spectra
by \citet{espaillat10} confirmed that the inner disk of this object
contained some optically thin dust and millimeter imaging by
\citet{hughes09} confirmed the size of the hole.  Here we take the model
presented in these previous works and modify it to better fit the
silicate dust features and the observed variability.
Figure~\ref{figgmaur} (top left) shows the three {\it Spitzer} IRS
spectra assembled for this study.  The GO1 and GO2 spectra agree with
one another.  GM~Aur displays seesaw-like behavior and a change in the
strength of the 10~{\micron} silicate emission.  However, the percentage
change in the flux at shorter wavelengths is larger than that seen at
the longer wavelengths beyond 20~{\micron}.

Eight additional SL spectra of GM~Aur were taken in GO Programs
20755 and 30896 (PI: Bary) in Oct 2005, March 2006, Oct 2006, and March 2007.  
We find no significant variability
between these spectra
and our GO spectra.  We note that in many of the spectra the flux of
the 10~{\micron} silicate
feature is $\sim$3--5$\%$ lower than in our GO observations and in AOR
19483648 the spectrum
is lower at all wavelengths by $\sim$9$\%$.  However, these small flux
changes are within the
mispointing uncertainties discussed in {\S}~\ref{mispoint}.  The
agreement between our GO spectra
and the spectra in Programs 20755 and 30896 indicates that the
change in GM~Aur's emission
occurred between the time the GTO observations were taken (27 February
2004) and the first epoch of Program 20755 (13 October 2005).
Two high resolution spectra were taken in
GO Program 30300 (PI: Najita) on March 14, 2007 using SH and LH.  One additional LH spectrum was taken in GTO Program 50498 (PI: Houck) on November 10, 2008.  
The spectra from GO Program 30300 agree with our GO observations between 10 and 15~{\micron}, 
but the observational uncertainties make it consistent with
both our GO and GTO observations at longer wavelengths.
The LH spectrum agrees with our GTO observations of GM Aur.  We do not have simultaneous data
at shorter wavelengths.  However, assuming that the seesaw-like behavior of GM Aur holds, one can conjecture that the flux in GM Aur changed within one month between our GO observations (taken in October 2008) back
to the flux observed in our GTO observations and the LH GTO observations.  More observations with {\it JWST} would be needed to 
test this.

The silicate dust composition used in the outer wall and disk is listed
in Table~\ref{tab:silwall} (Figure~\ref{figgmaur}, top left; GTO fit:
solid line; GO1$/$GO2 fit: broken line).  The hole is 23~AU in radius. 
The optically thin dust region extends up to 1~AU in the hole and
contains 2$\times$10$^{-12}$ M$_{\sun}$ of 0.005--1~{\micron} dust
composed of 32$\%$ organics, 12$\%$ troilite, and 56$\%$ silicates (see
Table~\ref{tab:silthin}).     We can reproduce the observed variability
by decreasing the amount of dust in the optically thin region between
the GTO and GO spectra by a factor of two.  Assuming an outer wall that
is not being shadowed, the outer wall changes in height from
2.9~AU in the GTO spectrum to 3.2~AU in the GO spectra.  This could
point to shadowing by the dust in the hole on the outer disk wall,
possibly due to opaque structure in the dust within the hole.  We do not
include the shadowing of this region and leave that to future work. In
the top right panel of Figure~\ref{figgmaur} we present the best fit
model (Table~\ref{tab:stellar}) to the broad-band SED.

\subsubsection{CS~Cha} \label{model:cscha}

CS~Cha was previously modeled by \citet{espaillat07a}.  It has been
identified as a spectroscopic binary \citep{guenther07}, but here we use
a circumstellar disk model for simplicity. Figure~\ref{figcscha}
(second, left) shows the three {\it Spitzer} IRS spectra assembled for
this study.  The GTO, GO1, and GO2 spectra all agree with one another
within the observational uncertainties except for the region around the
10~{\micron} silicate emission feature.
Two  additional IRS spectra of CS~Cha were taken in GO Program
30300 (PI: Najita) on August 2, 2006.  
This spectrum was taken using
the SH and LH modules.   We find that the SH spectrum
agrees with our GTO data (taken in July 2005) at ${\lambda}<$13~{\micron}
within the observational uncertainties.   The SH spectrum is consistent
with both our GTO and GO observations at ${\lambda}>$13~{\micron}.  The LH spectrum
is consistent with our GTO and GO observations.

We found that we could fit the IRS spectrum well with a 38~AU hole and a
silicate dust mixture in the outer wall and disk listed in
Table~\ref{tab:silwall} (Figure~\ref{figcscha}, bottom left; GTO fit:
solid line; GO1$/$GO2 fit: broken line).  The optically thin dust region
extends up to 1~AU in the hole and contains 1$\times$10$^{-12}$ M$_{\sun}$ 
of 0.005--4~{\micron} dust composed of
15$\%$ organics, 11$\%$ troilite, and 74$\%$ silicates (see
Table~\ref{tab:silthin}).  We note that the size and location of the
optically thin region was calculated without taking into account at
which radii dust can exist unperturbed in a binary system.  However,
lacking near-IR interferometry and sufficient details on the orbit of
the CS Cha system, we do not pursue this further.

The only variability observed occurs around the 10~{\micron} silicate
emission feature.  Relative to the time the GTO spectrum was taken, this
silicate emission has increased in the GO spectra.  This can be
explained by increasing the amount of dust in the optically thin inner
region by 17$\%$.  In Figure~\ref{figcscha} (bottom, right
panel) we present the best fit model to the broad-band SED. Disk
properties can be found in Table~\ref{tab:disk}.

\subsubsection{DM~Tau} \label{model:dmtau}

DM~Tau was previously modeled in detail by \citet{calvet05}.  They found
it had an inner disk hole of 3~AU that was empty of small dust grains. 
Veiling analysis using near-infrared spectra by \citet{espaillat10}
confirmed that the inner disk of this object was devoid of sub-micron or
micron-sized dust grains.  Here we take the model presented in this
previous work and modify it to better fit the silicate dust features in
the SED and the observed variability. Figure~\ref{figdmtau} (top, left)
shows the three {\it Spitzer} IRS spectra assembled for this study.  The
GTO, GO1, and GO2 spectra all agree with one another.

Eight additional SL spectra of DM~Tau were taken in GO Programs 
20755 and 30896 (PI: Bary) in September 12, 2005, October 2005, March 2006, October 2006, and March 2007.  Two additional high resolution spectra were taken in
GO Program 30300 (PI: Najita) on March 14, 2006 using SH and LH.  One additional LH spectra was taken by GTO Program 50498 (PI: Houck) on
November 10, 2008.  We find no variability between these spectra
and those in this paper within the observational uncertainties.  We note
that several campaigns in Program 20755 suffered from mispointing.  Since
no LL spectra were taken, we cannot discern the degree to which the spectra were
affected.  Therefore, we do not consider the decrease in flux in the mispointed 
observations to be intrinsic to the source.

We fit the IRS spectrum with a 3~AU hole and the silicate dust mixture
listed in Table~\ref{tab:silwall} (Figure~\ref{figdmtau}, top left,
solid line).  There is no optically thin dust within the gap.  This may
be the reason why no variability is observed in this object.  In the top right panel of
Figure~\ref{figdmtau} we present the best fit model to the broad-band
SED. Disk model properties are listed in Table~\ref{tab:disk}.

\subsubsection{T25}  \label{model:t25}

\begin{figure}
\epsscale{1}
\plotone{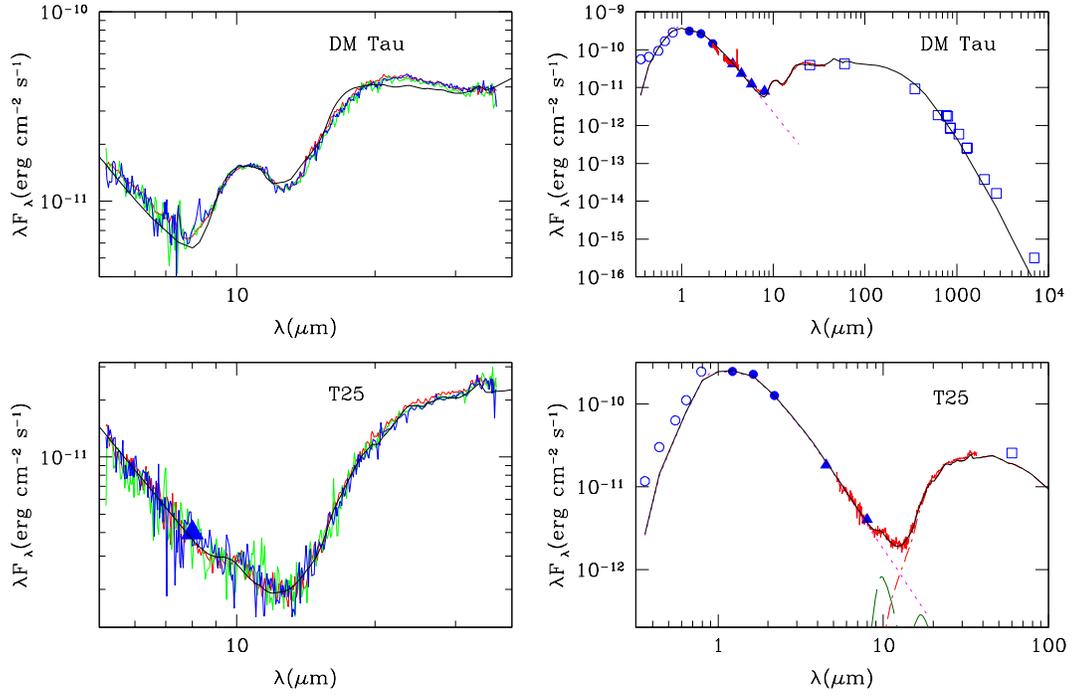}
\vskip -2in
\caption[]{SEDs and disk models of DM~Tau and T25.  
The left panels show the IRS GTO (red), GO1 (green), and GO2 (blue)
spectra for the target plotted with the best fit models (solid and
broken black lines; refer to the Appendix for more details).
Refer to the caption of 
Figure~\ref{figlkca15} for more details on the plotted data.  SpeX spectra first presented 
in \citet{espaillat10} are also plotted for DM Tau.
Millimeter data for DM Tau are taken from \citet{andrews05,
andrews07, rodmann06, kitamura02, dutrey96, beckwith90, beckwith91}.
(A color version of this figure is available in the online journal.)
}
\label{figdmtau}
\label{figsz18}
\end{figure}

\begin{figure}
\epsscale{1}
\plotone{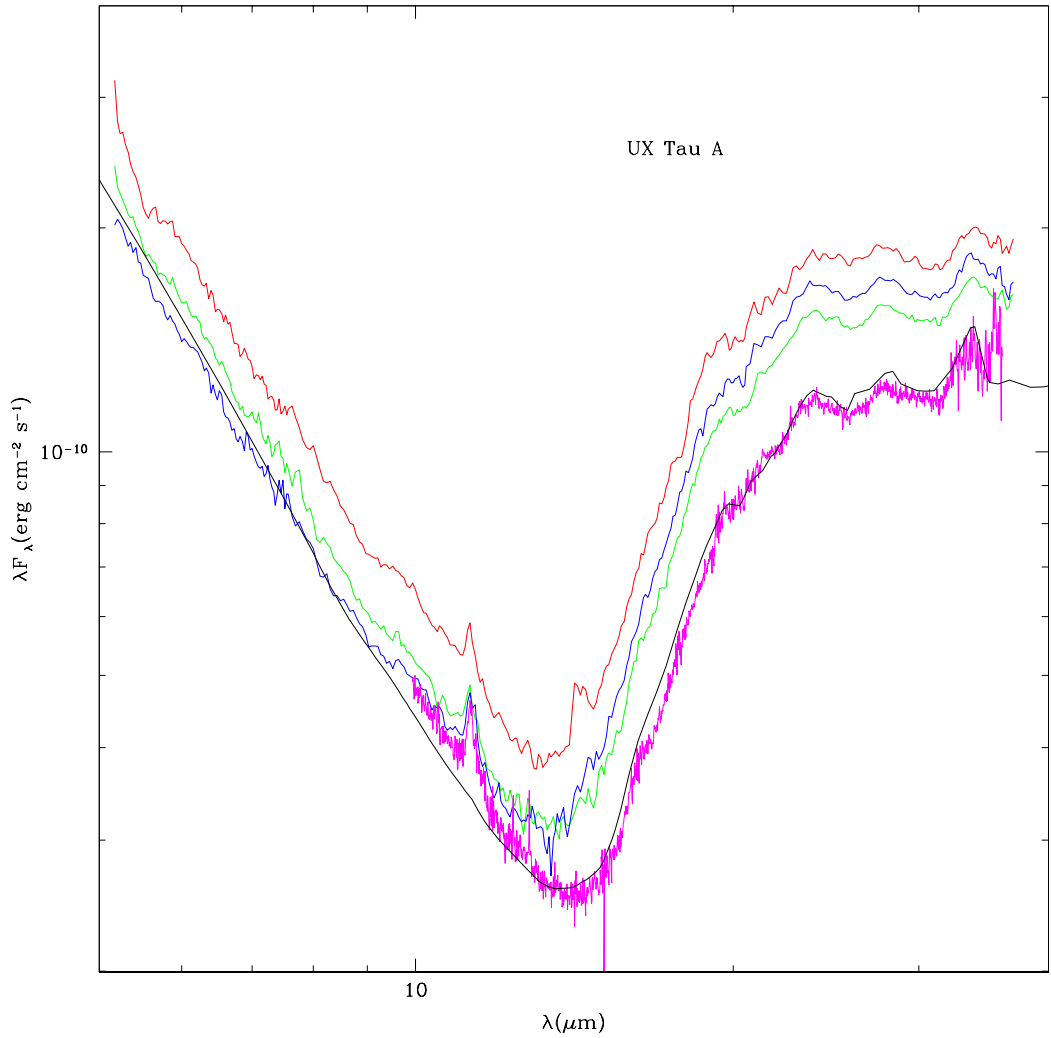}
\caption[]{SHLH spectrum and disk model fit for UX~Tau~A. 
Here we show the IRS GTO (red), GO1 (green), and GO2 (blue)
spectra for UX Tau A from Figure~\ref{figlkca15} plotted with the unbinned SHLH spectrum (magenta). The best fit model to the SHLH spectrum is shown (solid black line; refer to the Appendix for more details).
See Figure~\ref{figlkca15} for model fits to the GO1 and GO2 spectra.
(A color version of this figure is available in the online journal.)
}
\label{uxshlh}
\end{figure}

T25 is a transitional disk in the Chamaeleon cloud, previously
identified by \citet{kim09}.  Here we present the first detailed
modeling of this object. Figure~\ref{figsz18} (bottom left) shows the
three {\it Spitzer} IRS spectra assembled for this study.  The GTO, GO1,
and GO2 spectra all agree with one another.
Two additional IRS spectra of T25 were taken in GO Program
30300 (PI: Najita) on June 21, 2007 with the
SH and LH modules.   We find no variability between these spectra
and those in this paper within the observational uncertainties.  

Using the silicate dust mixture listed in Table~\ref{tab:silwall} and  a
hole of 13~AU, we modeled the contribution to the SED from the outer
wall and disk (Figure~\ref{figsz18}, bottom left, solid line).  The
optically thin dust region extends up to 1~AU in the hole and contains
ISM-sized dust composed of 19$\%$ organics, 15$\%$ troilite, and 66$\%$
silicates.  The silicates are made up of olivines. The gap contains
1$\times$10$^{-13}$ M$_{\sun}$ of small optically thin dust, the
smallest amount of such dust seen in this sample.  This reflects the
fact that the 10~{\micron} silicate emission feature is very weak.  The
SNR in this region is also very low and follow-up observations probing
if there is indeed 10~{\micron} silicate emission in this object would
be useful.  The combination of the weak silicate emission feature and poor
SNR at these wavelengths makes our derivation of the silicate composition in the optically
thin inner region more uncertain.

In the middle right panel of Figure~\ref{figsz18} we present the best
fit model to the broad-band SED.  Derived disk parameters are listed in
Table~\ref{tab:disk}. We could only find a millimeter upper limit from
\citet{henning93} which is not useful in constraining the outer disk. 
In this object, mostly all of the excess emission above the stellar photosphere
in the IRS wavelength range can be attributed to the outer wall.

\clearpage

\bibliographystyle{apjv2}

\clearpage

\begin{deluxetable}{lllcl}
\tabletypesize{\scriptsize}
\tablewidth{0pt}
\tablecaption{Log of Observations \label{tab:log}}
\startdata
\hline
\hline
Target & Alt. Names & Label  & Date & AOR ID\\
\hline
CR Cha		&  T 8 & GTO & 2005-05-26  &  12697345\\
...      		   	 & SZ 6 & GO1 & 2008-06-01  &  26143744\\
...                    	& CHX 3 & GO2 & 2008-06-08 & 27186944 \\
CS~Cha		& T 11 &  GTO & 2005-07-11  & 12695808\\
...      		   	 &  & GO1 & 2008-06-01  &  26144000\\
...                    	& & GO2 & 2008-06-08 & 27187712 \\
DM~Tau		&  & GTO1 & 2004-02-08  & 3536384\\
...      		   	 &  &GTO2 & 2006-03-15  &  16346624\\
...      		   	 & & GO1 & 2008-10-01  &  26141952\\
...                    	&  & GO2 & 2008-10-08 &  27184640\\
GM~Aur		&  & GTO & 2004-02-27  & 3538944\\
...      		   	 & & GO1 & 2008-10-07  & 26141696 \\
...                    	& & GO2 & 2008-10-14 &  27186688\\
IP~Tau		&  & GTO & 2004-02-08  & 3535616\\
...      		   	 & & GO1 & 2008-10-02  & 26141440 \\
...                    	& & GO2 & 2008-10-08 & 27184896 \\
ISO 52			& B18 & GTO & 2005-07-12  &  12691200\\
...      		   	 &  & GO1 & 2008-06-01  &  26144768\\
...                    	&  & GO2 & 2008-06-08 & 27187200 \\
LkCa~15		&  & GTO & 2004-02-08  & 3537664\\
...      		   	 & & GO1 & 2008-10-02  &  26140672\\
...                    	& & GO2 & 2008-10-08 &  27186176\\
RY~Tau		&  & GTO$^{a}$ & 2004-03-02  & 3529984\\
...      		   	 & & GO1$^{b}$ & 2008-10-07  & 26141184 \\
...                    	& & GO2$^{b}$ & 2008-10-14 &  27185920 \\
SZ~Cha		&  T6 & GTO & 2005-07-12  & 12696832\\
...      		   	 & & GO1 & 2008-06-01  & 26142464 \\
...                    	& & GO2 & 2008-06-08 & 27187968 \\
T25		& SZ 18 &  GTO & 2005-04-24  & 1269552\\
...      		   	 & & GO1 & 2008-06-01  &  26144256\\
...                    	& & GO2 & 2008-06-08 & 27185152 \\
T35		&  SZ 27 & GTO & 2005-07-13  & 12696320 \\
...      		   	 &  & GO1 & 2008-06-01  &  26143488\\
...                    	& & GO2 & 2008-06-08 &  27185664\\
T56		& SZ 45 & GTO & 2005-07-13  & 12696064 \\
...      		   	 & HM 32 & GO1 & 2008-06-01  & 26142720 \\
...                    	& & GO2 & 2008-06-08 &  27186432\\
UX~Tau~A		&  & GTO & 2004-02-08  & 3536384\\
...      		   	 & & GO1 & 2008-10-01  &  26140928\\
...                    	& & GO2 & 2008-10-08 & 27187456 \\
WW~Cha		&  T44& GTO & 2006-03-07  & 12697088\\
...      		   	 & SZ 34  & GO1$^{a}$ & 2008-06-01  &  26142976\\
...                    	& & GO2$^{a}$ & 2008-06-09 & 27185408
\tablenotetext{a}{These observations were slightly mispointed.}
\tablenotetext{b}{These spectra were saturated in the 10~{\micron} region.}

\enddata
\end{deluxetable}

\begin{deluxetable}{lcccccccc}
\tabletypesize{\scriptsize}
\tablewidth{0pt}
\tablecaption{Stellar Properties of Sample\label{tab:stellar}}
\startdata
\hline
\hline
Target & A$_V$ & Spectral  & T$_{*}$ &  L$_{*}$          &M$_{*}$         & R$_{*}$          &  $\mdot$                       & Distance \\
             &               &Type        &(K)           & (M$_{\sun}$) & (M$_{\sun}$) & (R$_{\sun}$) &  (10$^{-9}$ M$_{\sun}$ yr$^{-1}$) & (pc)   \\
\hline
CR~Cha		&  1.5 & K2$^{b}$   	& 4900 & 3.5 & 1.9  & 2.6  & 8.8 & 160 \\
CS~Cha		&  0.8 & K6$^{a}$      	& 4205 & 1.5 & 0.9  & 2.3  & 12  & 160 \\
DM~Tau		&  0.7 & M1.5$^{c}$  	& 3720 & 0.3  & 0.5  & 1.3 & 3.1 & 140 \\
GM~Aur		&  0.8 & K5.5$^{c}$  	& 4350 & 0.9  & 1.1  & 1.7 & 4.7 & 140 \\
IP~Tau		&  0.5 & M0$^{d}$  	& 3850 & 0.5  & 0.6  & 1.6 & 0.42 & 140 \\
ISO~52			&  1.3 & M4$^{a}$     	& 3370 & 0.1 & 0.3  & 1.0  & --  & 160 \\
LkCa~15		&  1.7 & K3$^{c}$  	& 4730 & 1.2  & 1.3  & 1.6 & 3.3 & 140 \\
RY~Tau		&  2.2 & G1$^{e}$ 	& 5945 & 11.7  & 2.2  & 3.2 & 9$^{e}$1 & 140 \\
SZ~Cha		&  1.9 & K0$^{b}$	& 5250 & 1.9  & 1.4  & 1.7 & 2.4 & 160 \\
T25		&  1.6 & M2.5$^{a}$   	& 3470 & 0.3 & 0.3  & 1.5  & 0.97  & 160 \\
T35		&  3.5 & M0$^{f}$  	& 3850 & 0.4 & 0.6  & 1.5  & 1.2  & 160 \\
T56		&  0.6 & M0.5$^{a}$   	& 3720 & 0.4 & 0.5  & 1.6  & 1.5  & 160 \\
UX~Tau~A		&  1.8 & G8$^{c}$  	& 5520 & 2.6 & 1.5  & 1.8  & 11  & 140 \\
WW~Cha		&  4.8 & K5$^{a}$   	& 4350 & 6.5 & 1.2  & 4.5  & --  & 160
\enddata
\tablenotetext{a}{\citet{luhman04}}
\tablenotetext{b}{\citet{rydgren80}}
\tablenotetext{c}{\citet{espaillat10}}
\tablenotetext{d}{\citet{herbig86}}
\tablenotetext{e}{\citet{calvet04}}
\tablenotetext{f}{\citet{gauvin92}}
\end{deluxetable}

\begin{deluxetable}{lccccccccccccccc}
\tabletypesize{\scriptsize}
\tablewidth{0pt}
\tablecaption{Disk Properties of Sample\label{tab:disk}}
\startdata
\hline
\hline
\multicolumn{1}{c}{Target} & \multicolumn{1}{c}{$i$} & \multicolumn{1}{c}{} & \multicolumn{4}{c}{Inner Wall$^{a}$} & \multicolumn{1}{c}{}& \multicolumn{4}{c}{Outer Wall$^{a}$} & \multicolumn{1}{c}{} & \multicolumn{3}{c}{Outer Disk}\\
\cline{4-7} 
\cline{9-12}
\cline{14-16}
     &  &  & a$_{max}$  & T$_{wall}^i$ &z$_{wall}^i$ & R$_{wall}^i$   & & a$_{max}$  & T$_{wall}^o$ &z$_{wall}^o$         & R$_{wall}^o$ & & $\epsilon$ &$\alpha$ & M$_{disk}$ \\
     &  &  & ({\micron})   &(K)                  &(AU)               & (AU)                  & & ({\micron})   &(K)                    &(AU)                        & (AU)                 &  &                   &                &(M$_{\sun}$)\\
\hline
CR~Cha		& 60             & & 1 & ... & ... & ... & & 0.25	& 220	& ... & 10 & & 0.01 & 0.0013&  0.1  \\
CS~Cha		& 60             & & -- & -- & -- & -- & & 4	& 90	& 7 & 38 & & 0.01 & 0.001	&  0.3  \\
DM~Tau		& 45$^{b}$ & & -- & -- & -- & -- & & 1	& 200	& 0.3 & 3 & & 0.5 & 0.001	&  0.1  \\
GM~Aur		& 64$^{c}$ & &	-- & -- & --	 & --	&	& 3	& 110	& 2.9(3.2) 	& 23	& 	& 0.01	 & 0.0004 & 0.4  \\
IP~Tau		& 60		  & & 10 & 1400	& 0.006(0.007) & 0.07 & & 4	& 230	& 36	& 2 	& & 0.01 & 0.0008	& 0.01	  \\
LkCa~15		& 42$^{b}$ & & 1 & 1400	& 0.009(0.007) 	& 0.15	& & 0.25 & 110 & 5 & 39 &  & 0.001& 0.0005 & 0.1\\
RY~Tau		& 71$^{d}$ & & 10 & 1400	& 0.02	& 0.42  &	& 3 & 220	& 4 & 18 &  & 0.01 & 0.01 &	0.2  \\
SZ~Cha		& 60 	 	 & & 5 & 1400 & 0.006(0.008)	& 0.16	& & 4	& 135 & 4 & 18  &  & 0.01 & 0.001 & 0.05 \\
T25		& 60            & & -- &  -- &  -- & -- 	&  & 5	& 100	& 2 & 13	& 	&  --	&  --	&  --	  \\
T35		& 60 		 & & 10 & 1800 & 0.005(0.006)	& 0.04	& & 3 & 100 & 4 & 15 &  & -- & -- & -- \\
T56		& 60 	 	 & & 10  & 1400 & 0.002(0.001) & 0.06	& & 0.25 & 100 & 4 & 20 & & 0.001	& 0.0001	& 0.3	 \\
UX~Tau~A		& 60   	 & & 10 & 1550	& 0.006(0.005)	& 0.15 & & 5	& 120 & 3 & 30 &  & 0.001 & 0.004 &~~~~~~~~~~~~0.05 

\tablenotetext{a}{For the wall heights, the values given correspond to the best fit to the GTO spectrum. The values relevant to the GO1 spectra are in parenthesis. For CR~Cha, see the Appendix for the values of the inner wall
temperature, height, and radius and the height of the outer wall.  In the case of UX~Tau~A, we do not fit the GTO spectrum.  Instead we list the height of the inner wall for the GO1 spectrum; the value for the GO2 spectrum is in parenthesis.  All other parameter values listed are the same between all the epochs.}
\tablenotetext{b}{\citet{simon00}}
\tablenotetext{c}{\citet{hughes09}}
\tablenotetext{d}{\citet{isella10}}
\enddata

\end{deluxetable}

\begin{deluxetable}{lccccc}
\tabletypesize{\scriptsize}
\tablewidth{0pt}
\tablecaption{Mass Fraction (in $\%$) of Silicates in Outer Wall \& Disk \label{tab:silwall}}
\startdata
\hline
\hline
Target & Amorphous  & Amorphous & Crystalline  &  Crystalline   & Crystalline  \\
            & Olivine           &  Pyroxene     &        Forsterite           &     Enstatite              & Silica  \\
 \hline
CR~Cha		& --  & 90 & -- & -- & 10 \\
CS~Cha		& --  & 98 & 2 & -- & -- \\
DM~Tau		& 65   & 30 & 2 & -- & 3 \\
GM~Aur		& 45  & 50 & -- & -- & 5 \\
IP~Tau		& 50  & 45 & 5 & -- & -- \\
ISO~52		& 80  & -- & 10 & 10 & -- \\
LkCa~15		&  88 & -- & 1 & -- & 11 \\
RY~Tau		& 100   & -- & -- & --  &-- \\
SZ~Cha		& 32   & 55 & 5 & 1 & 7 \\
T25			& 90  & -- & 10 & -- & -- \\
T35			& 40   & -- & 30 & 30 & -- \\
T56			& --  & 82 & 8 & -- & 10 \\
UX~Tau~A	& 81  & -- & 10 & 4 & 5
\enddata
\end{deluxetable}

\begin{deluxetable}{lccccc}
\tabletypesize{\scriptsize}
\tablewidth{0pt}
\tablecaption{Mass Fraction (in $\%$) of Silicates in Optically Thin Dust Region \label{tab:silthin}}
\startdata
\hline
\hline
Target & Amorphous  & Amorphous & Crystalline  &  Crystalline   & Crystalline  \\
            & Olivine           &  Pyroxene     &        Forsterite           &     Enstatite              & Silica  \\
 \hline
CR~Cha		& --  & 90 & -- & -- & 10 \\
CS~Cha		& 95  & -- & -- & 5 & -- \\
GM~Aur		&  90  & -- & 10 & -- & -- \\
IP~Tau		& 50  & 45 & 5 & -- & -- \\
ISO~52		& 80  & -- & 10 & 10 & -- \\
LkCa~15$^{a}$ &  90 & -- & 10 & -- & -- \\
LkCa~15$^{b}$ &  -- & 100 & -- & -- & -- \\
RY~Tau		& 100   & -- & -- & --  &-- \\
SZ~Cha		& 90  & -- & 10 & -- & -- \\
T25		& 100   & -- &  -- & -- & -- \\
T56		& 41  & 34 & 10 & 15 & --
\enddata
\tablenotetext{b}{Values are for optically thin dust located within 1~AU of the gap.}
\tablenotetext{c}{Values are for optically thin dust located between 1 and 15~AU of the gap.}
\end{deluxetable}

\begin{deluxetable}{lccc}
\tabletypesize{\scriptsize}
\tablewidth{0pt}
\tablecaption{Spitzer Photometry\label{tab:spitzer}}
\startdata
\hline
\hline
Target & Instrument$^{a}$ & Date  & Symbol$^{b}$ \\
\hline
CR~Cha		& -- & -- & --  \\
CS~Cha		& IRAC & 01 May 2004  &  \\
...			& IRAC & 04 Jul 2004  & open blue\\
...			& IRAC & 10 Jul 2006  & open red\\
...			& MIPS 24~{\micron} & 08 Apr 2004 &  \\
...			& MIPS 24~{\micron} & 22 Jun 2004 & open blue\\
...			& MIPS 24~{\micron} & 27 Feb 2005 & open red \\
DM~Tau		& IRAC & 19 Feb 2005  &   \\
GM~Aur		& IRAC & 14 Feb 2004  &  \\
...			& MIPS 24~{\micron} & 05 Mar 2005 &   \\
IP~Tau		& IRAC & 20 Feb 2005  & \\
...			& IRAC & 21 Feb 2005  & open blue \\
...			& IRAC & 16 Oct 2007  & open red \\
...			& MIPS 70~{\micron} & 27 Feb 2005 & \\
ISO~52		& IRAC & 19 Feb 2004  &   \\
...			& IRAC & 10 Jun 2004  & open red \\
...			& IRAC & 09 Jul 2004  & open green \\
...			& IRAC & 02 Sep 2004  & open magenta \\
...			& MIPS 24~{\micron} & 08 Apr 2004 & \\
...			& MIPS 24~{\micron} & 27 Feb 2005 & open red \\
LkCa~15		& IRAC & 20 Feb 2005  &    \\
...			& IRAC & 3 Apr 3 2007  & open blue  \\
...			& MIPS 24~{\micron} & 26 Feb 2007 & \\
...			& MIPS 70~{\micron} & 27 Feb 2007 & \\
RY~Tau		& IRAC & 20 Feb 2005  &    \\
...			& IRAC & 21 Feb 2005  & open blue \\
SZ~Cha		& IRAC & 10 Jul 2006  &  \\
T25			& IRAC & 10 Jul 2006  &   \\
T35			& IRAC & 10 Jun 2004  &    \\
...			& MIPS 24~{\micron} & 27 Feb 2005 & \\
T56			& -- & -- & --  \\
UX~Tau~A	& IRAC & 19 Feb 2005   & \\
...			& MIPS 24~{\micron} & 25 Sept 2004 & \\
WW~Cha		& IRAC & 04 Jul 2004  &
\enddata
\tablenotetext{a}{IRAC and MIPS 24~{\micron} photometry for objects in Taurus are from \citet{luhman10}; for  Chamaeleon objects, these data are from \citet{luhman08}.
All MIPS 70~{\micron} photometry are taken from \citet{rebull10}}
\tablenotetext{b}{All IRAC and MIPS photometry are plotted as closed, blue triangles unless otherwise noted.}
\end{deluxetable}

\clearpage

\clearpage

\clearpage

\clearpage

\clearpage

\clearpage

\clearpage

\clearpage

\clearpage

\clearpage

\clearpage

\clearpage

\clearpage

\clearpage

\clearpage

\clearpage

\clearpage

\clearpage

\end{document}